\documentclass[12pt]{article}
\usepackage{graphicx}

\begin{document}
\title{Model independent search for $Z'$-boson signals}
\author{A. V. Gulov, V. V. Skalozub\\[3mm]
\it Dnipropetrovsk National University, Dnipropetrovsk, Ukraine}
\date{\empty}
\maketitle
\begin{abstract}
An approach to the model-independent searching for the $Z'$ gauge
boson  as a virtual state in scattering processes is developed. It
accounts for as a basic requirement the renormalizability of
underlying unspecified in other respects model. This  results in a
set of relations between low energy couplings of $Z'$ to fermions
that reduces in an essential way the number of parameters to be
fitted in experiments. On this ground the observables which
uniquely pick out the $Z'$ boson in leptonic processes are
introduced and the data of LEP  experiments analyzed. The $Z'$
couplings to leptons and quarks are estimated at 95\% confidence
level. These estimates may serve as a guide for experiments at the
Tevatron and/or LHC. A comparison with other approaches and
results is given.
\end{abstract}
\thispagestyle{empty}

\newpage
\tableofcontents

\newpage
\section{Introduction}
The precision test of the standard model (SM) at the LEP gave a
possibility not only to determine all the parameters and particle
masses at the level of radiative corrections but also afforded an
opportunity for searching for signals of new heavy particles
beyond the energy scale of it. On the base of the LEP2 experiments
the low bounds on parameters of various models extending the SM
have been estimated and the scale of new physics was obtained
\cite{EWWG,OPAL,DELPHI}. Although no new particles were
discovered, a general believe is that the energy scale of new
physics to be of order 1 TeV, that may serve as a guide for
experiments at the Tevatron and  LHC. In this situation, any
information about new heavy particles obtained on the base of the
present day data is desirable and important.

Numerous extended models include the $Z'$ gauge boson -- massive
neutral vector particle associated with the extra $U(1)$ subgroup
of an underlying group. Searching for this particle as a virtual
state is widely discussed in the literature (see for references
\cite{Leike,Lang08}). In the content of searching for $Z'$ at the
LHC and the ILC an essential information and prospects for future
investigations  are given in lectures \cite{Rizzo06}. Such aspects
as the mass of $Z'$, couplings to the SM particles, $Z - Z'$
mixing and its influence in various processes and particles
parameters, distinctions between different models are discussed in
details. We shall turn to these papers in what follows. As
concerned a searching for $Z'$ in the LEP experiments and the
experiments at Tevatron \cite{Ferroglia}, it was carried out
mainly in a model-dependent way. A wide class of popular models
has been investigated and  low bounds on the mass $m_{Z'}$ were
estimated (see, \cite{EWWG,OPAL,DELPHI}). As it is occurred, the
low masses are varying in a wide energy interval 400-1800 GeV
dependently on a specific model. These bounds are a little bit
different in the LEP and Tevatron  experiments. In this situation
a model-independent analysis is of interest.

In the papers \cite{EPJC2000,PRD2000,YAF2004,PRD2004} of the
present authors a new approach for the model-independent search
for $Z'$-boson was proposed which, in contrast to other
model-independent searches, gives a possibility to pick out
uniquely this virtual state and determine its characteristics. The
corresponding observables have also been introduced and applied to
analyze the LEP2 experiment data. Our consideration is based on
two constituents: 1) The relations between the effective
low-energy couplings derived from the renormalization group (RG)
equation for fermion scattering amplitudes. We called them the RG
relations. Due to these relations, a number of unknown $Z'$
parameters entering the amplitudes of different scattering
processes considerably decreases. 2) When these relations are
accounted for, some kinematics properties of the amplitude become
uniquely correlated with this virtual state and the $Z'$ signals
exhibit themselves.

The RG relations allow to introduce observables correlated
uniquely with the $Z'$-boson. Comparing the mean values of the
observables with the necessary specific values, one could arrive
at a conclusion about the $Z'$ existence. The confidence level
(CL) of these values has been estimated and adduced in addition.
Without taking into consideration the RG relations the
determination of $Z'$-boson requires a supplementary specification
due to a larger number of different couplings contributing to the
observables. Similar situation takes place in the ``helicity model
fits'' of LEP Collaborations \cite{EWWG,OPAL,DELPHI} when
different virtual states contribute to each of the specific models
(AA, VV, and so on). Therefore these fits had the goal to discover
any signals of new physics independently of the particular states
which may cause deviations from the SM. Note that the LEP
Collaborations saw no indications of new contact four fermion
interactions in these fits.

In Refs. \cite{EPJC2000,YAF2004,PRD2004} the one-parametric
observables were introduced and the signals (hints in fact) of the
$Z'$ have been determined at the 1$\sigma$ CL in the
$e^+e^-\to\mu^+\mu^-$ process, and at the 2$\sigma$ CL in the
Bhabha process. The $Z'$ mass was estimated to be 1--1.2 TeV. An
increase in statistics could make these signals more pronounced
and there is a good chance to discover this particle at the LHC.

In Ref. \cite{PRD2007} the updated results of the one-parameter
fit and  the complete many-parametric fit of the LEP2 data were
performed  with the goal  to estimate a possible signal of the
$Z'$-boson with accounting for the final data of the LEP
collaborations DELPHI and OPAL \cite{OPAL,DELPHI}. Usually, in a
many-parametric fit the uncertainty of the result increases
drastically because of extra parameters. On the contrary, in our
approach due to the RG relations between the low-energy couplings
there are only 2-3 independent parameters for the investigated
leptonic  scattering processes. As it was  showed in Ref.
\cite{PRD2007}, an inevitable increase of confidence areas in the
many-parametric space was compensated due to accounting for all
accessible experimental information. Therefore, the uncertainty of
the many-parametric fit was estimated as the comparable with
previous one-parametric fits in Refs. \cite{YAF2004,PRD2004}. In
this approach the combined data fit for all lepton processes is
also possible.

From the results obtained on the  searching for Abelian $Z'$
within the LEP experiment data set we conclude that it is
insufficient for convincing discovery of this particle as the
virtual state. In this situation it is reasonable to  analyze  the
data by using the neural network approach which is able to make a
realistic prognoses for the parameters of interest. This
investigation was done within the two parametric global fit of the
LEP2 data on the Bhabha scattering process. As the result of all
these considerations we derive at the 2$\sigma$ CL the
characteristics of the $Z'$ (the vector $v$ and axial-vector $a$
couplings of the $Z'$ with SM leptons and the $Z-Z'$ mixing). The
$Z'$ mass is also estimated. Due to the universality of the $a$ we
also derived the model independent estimate of the $Z'$
axial-vector couplings to quarks, $a_q = a$. Note that the hints
for the $Z'$ have been determined in all the processes considered
that increases the reliability of the signal. These  results may
serve as a good input into the future LHC and ILC experiments and
used in various aspects. To underline the importance of them we
mention that there are many tools at the LHC for the
identification  of $Z'$. But many of them are only applicable if
$Z'$ is relatively light. The knowledge of the $Z'$ couplings to
SM fermions also have important consequences.

The paper is organized as follows. In sect. 2 we give a necessary
information about the description of $Z'$ at low energies. In
sects. 3-5 we discuss the origin of the RG relations, their
explicit forms for the case   of the heavy $Z'$ and consequences
of the relations for scattering processes investigated. In sect. 6
the cross sections and the observables to pick out uniquely the
virtual $Z'$ in the $e^+ e^- \to \mu^+ \mu^-, \tau^+ \tau^-$
processes  are given. The fits of data are described and
discussed. Then in sect. 7 the same is present for the Bhabha
process $e^+ e^- \to e^+ e^-$. The one parametric and two
parametric fits are discussed. In sect. 8 the analysis of this
process is carried out  by using the neuron network approach. The
criteria for training the network are introduced which guarantee
the 2$\sigma$ CL deviations of the data from the model containing
the SM with extra $ Z'$. The obtained parameters of the $Z'$
practically coincide with that of derived in the one parameter
analysis. In this way we determine the characteristics of the $Z'$
coming from the LEP experiments. In sect. 9 we discuss the role of
the present model-independent analysis  for the LHC experiments.
The discussion and comparison with results of other approaches are
given in sect. 10. In the Appendix we describe the two-mass-scale
Yukawa model and analyze in detail how the decoupling of the loop
contributions due to heavy virtual states is realized when the
mixing of fields is taken into consideration. This point is an
essential element of the approach developed.

\section{The Abelian $Z'$ boson at low energies}

Let us adduce a necessary information about the Abelian
$Z'$-boson. This particle is predicted by a number of grand
unification models. Among them the $E_6$ and $SO(10)$ based models
\cite{Hewett} (for instance, LR, $\chi-\psi$ and so on) are often
discussed in the literature. In all the models, the Abelian
$Z'$-boson is described by a low-energy $\tilde{U}(1)$ gauge
subgroup originated in some symmetry breaking pattern.

At low energies, the $Z'$-boson can manifest itself by means of
the couplings to the SM fermions and scalars as a virtual
intermediate state. Moreover, the $Z$-boson couplings are also
modified due to a $Z$--$Z'$ mixing. In principle, arbitrary
effective $Z'$ interactions to the SM fields could be considered
at low energies. However, the couplings of non-renormalizable
types have to be suppressed by heavy mass scales because of
decoupling. Therefore, significant signals beyond the SM can be
inspired by the couplings of renormalizable types. Such couplings
can be derived by adding new $\tilde{U}(1)$-terms to the
electroweak covariant derivatives $D^\mathrm{ew}$ in the
Lagrangian \cite{cvetic87,degrassi89} (review,
\cite{Leike,Lang08})
\begin{eqnarray} \label{Lf}
L_f &=& i \sum\limits_{f_L} \bar{f}_L \gamma^\mu \left(
\partial_\mu - \frac{i g}{2} \sigma_a W^a_\mu - \frac{i g'}{2}
B_\mu Y_{f_L} -
\frac{i \tilde{g}}{2}\tilde{B}_\mu \tilde{Y}_{f_L}\right) f_L \\
\nonumber &+& i \sum\limits_{f_R} \bar{f}_R \gamma^\mu \left(
\partial_\mu  - i
g' B_\mu Q_{f} - \frac{i \tilde{g}}{2}\tilde{B}_\mu
\tilde{Y}_{f_R}\right) f_R,
\end{eqnarray}
where summation over all the SM left-handed fermion doublets,
leptons and quarks, $f_L = {(f_u)_L, (f_d)_L}$, and the
right-handed singlets, $f_R = (f_u)_R, (f_d)_R $, is understood.
$Q_f$ denotes the charge of $f$ in positron charge units,
$\tilde{Y}_{f_L}=\mathrm{diag}(\tilde{Y}_{f_u}, \tilde{Y}_{f_d})$,
and $Y_{f_L}= -1$ for leptons and 1/3 for quarks.

For general purposes we derive the RG relations for the $Z'$
beyond the SM with two light Higgs doublets (THDM)
\cite{EPJC2000}. $Z'$ interactions with the scalar doublets can be
parametrized in a model-independent way as follows,
\begin{eqnarray} \label{Lscal}
L_\phi = \sum\limits_{i=1}^{2} \left| \left( \partial_\mu -
\frac{i g}{2} \sigma_a W^a_\mu - \frac{i g'}{2} B_\mu Y_{f_L}  -
\frac{i \tilde{g}}{2} \tilde{B}_\mu \tilde{Y}_{\phi_{i}} \right)
\phi_i\right|^2.
\end{eqnarray}
In these formulas, $g, g', \tilde{g}$ are the charges associated
with the $SU(2)_L, U(1)_Y,$ and  the $Z'$ gauge groups,
respectively, $\sigma_a$ are the Pauli matrices,
$\tilde{Y}_{\phi_{i}} = \mathrm{diag}(\tilde{Y}_{\phi_{i,1}},
\tilde{Y}_{\phi_{i,2}}) $ is the generator corresponding to the
gauge group of the $Z'$ boson, and $Y_{\phi_i}$ is the $U(1)_Y$
hypercharge.

The Yukawa Lagrangian can be written in the form
\begin{eqnarray} \label{Lyukawa}
L_\mathrm{Yuk.} &=& - \sqrt{2}
\sum\limits_{f_L}\sum\limits_{i=1}^{2} \left(G_ {f_{d,i}}
[\bar{f}_L \phi_i (f_d)_R + (\bar{f}_d)_R \phi^{+}_i f_L]
\right.\nonumber\\&&\left. +G_{f_{u,i}} [\bar{f}_L \phi_i^c
(f_u)_R + (\bar{f}_u)_R \phi^{c+}_i f_L]\right),
\end{eqnarray}
where $\phi^c_i = i \sigma_2 \phi^*_i $  is the charge conjugated
scalar doublet.

The Lagrangian (\ref{Lscal}) leads to the $Z$--$Z'$ mixing. The
$Z$--$Z'$ mixing angle $\theta_0$ is determined by the coupling
$\tilde{Y}_\phi$ as follows
\begin{equation}\label{2}
\theta_0 =
\frac{\tilde{g}\sin\theta_W\cos\theta_W}{\sqrt{4\pi\alpha_\mathrm{em}}}
\frac{m^2_Z}{m^2_{Z'}} \tilde{Y}_\phi
+O\left(\frac{m^4_Z}{m^4_{Z'}}\right),
\end{equation}
where $\theta_W$ is the SM Weinberg angle, and
$\alpha_\mathrm{em}$ is the electromagnetic fine structure
constant. Although the mixing angle is a small quantity of order
$m^{-2}_{Z'}$, it contributes to the $Z$-boson exchange amplitude
and cannot be neglected at the LEP energies.

In what follows we will also use the $Z'$ couplings to the vector
and axial-vector fermion currents defined as
 \begin{equation} \label{av} v_f =
\tilde{g}\frac{\tilde{Y}_{L,f} + \tilde{Y}_{R,f}}{2}, \qquad a_f =
\tilde{g}\frac{\tilde{Y}_{R,f} - \tilde{Y}_{L,f}}{2}.
\end{equation}
The Lagrangian (\ref{Lf}) leads to the following interactions
between the fermions and the $Z$ and $Z'$ mass eigenstates:
\begin{eqnarray}
{\cal L}_{Z\bar{f}f}&=&\frac{1}{2} Z_\mu\bar{f}\gamma^\mu\left[
(v^\mathrm{SM}_{fZ}+\gamma^5 a^\mathrm{SM}_{fZ})\cos\theta_0
+\right.\nonumber\\&&\quad\left.
+(v_f+\gamma^5 a_f)\sin\theta_0 \right]f, \nonumber\\
{\cal L}_{Z'\bar{f}f}&=&\frac{1}{2} Z'_\mu\bar{f}\gamma^\mu\left[
(v_f+\gamma^5 a_f)\cos\theta_0 -\right.\nonumber\\&&\quad\left.
-(v^\mathrm{SM}_{fZ}+\gamma^5
a^\mathrm{SM}_{fZ})\sin\theta_0\right]f,
\end{eqnarray}
where $f$ is an arbitrary SM fermion state; $v^\mathrm{SM}_{fZ}$,
$a^\mathrm{SM}_{fZ}$ are the SM couplings of the $Z$-boson.

Since the $Z'$ couplings enter the cross-section together with the
inverse $Z'$ mass, it is convenient to introduce the dimensionless
couplings
\begin{equation}\label{6}
\bar{a}_f=\frac{m_Z}{\sqrt{4\pi}m_{Z'}}a_f,\quad
\bar{v}_f=\frac{m_Z}{\sqrt{4\pi}m_{Z'}}v_f,
\end{equation}
which can be constrained by experiments.

Low energy parameters $\tilde{Y}_{\phi_{i,1}}$,
$\tilde{Y}_{\phi_{i,2}}$, $\tilde{Y}_{L,f}$, $\tilde{Y}_{R,f}$
must be fitted in experiments. In most investigations they were
considered as independent ones. In a particular model, the
couplings $ \tilde{Y}_{\phi_{i,1}}$, $ \tilde{Y}_{\phi_{i,2}},
\tilde{Y}_{L,f}, \tilde{Y}_{R,f}$ take some specific values. In
case when the model is unknown, these parameters  remain
potentially arbitrary numbers. However, this is not the case if
one assumes that the underlying extended model is a renormalizable
one. In the papers \cite{EPJC2000,PRD2000} it was shown that these
parameters are correlated due to renormalizability. We called them
the RG relations. Since this notion is a key-point of our
consideration, we discuss it in detail.

\section{Renormalization group relations}

What is RG relation? Generally speaking, this is a correlation
between  low energy parameters of interactions of a heavy new
particle with  known light particles of the SM following from the
requirement that full unknown yet theory extending SM is to be
renormalizable.

Strictly speaking, RG relations are the consequence of two
constituencies:
\begin{enumerate}
\item RG equation for a scattering amplitude;
\item Decoupling theorem.
\end{enumerate}

The latter one describes the modification of both  the RG operator
\begin{equation} \label{rgo}
{{\cal D}} = \frac{d}{d \log \mu} = \frac{\partial}{\partial \log
\mu} + \sum\limits_a \beta_a \frac{\partial}{\partial
\hat{\lambda}_a} - \sum\limits_{\hat{X}} \gamma_X
\frac{\partial}{\partial \log \hat{X}}
\end{equation}
and an amplitude at the energy threshold $\Lambda$ of new physics.
Here, $\beta_a$- and $\gamma_X$-functions correspond to all the
charges $\hat{\lambda}_a$ and fields and masses $\hat{X}$ of the
underlying theory.

The RG equation for a scattering amplitude $f$ reads,
\begin{equation} \label{RGe}
{{\cal D}}f  = \left(\frac{\partial}{\partial \log \mu} +
\sum\limits_a \beta_a \frac{\partial}{\partial \hat{\lambda}_a} -
\sum\limits_{\hat{X}} \gamma_X \frac{\partial}{\partial \log
\hat{X}}\right)f = 0,
\end{equation}
where $f$ accounts for as intermediate states either the light or
heavy virtual particles of the full theory.  The standard usage of
the RG equation is to improve the amplitude by solving this
equation for the operator ${{\cal D}}$ calculated in a given order
of perturbation theory. However, to search for heavy virtual
particles, we will use Eq. (\ref{RGe}) in another way.

First we note that for any  renormalizable theory, the RG equation
is just identity, if  $f$ and ${{\cal D}}$ are calculated in a
given order of loop expansion. In this case Eq. (\ref{RGe})
expresses the well known fact that the structure of the divergent
term coincides with the structure of the corresponding term in a
tree-level Lagrangian.

For example, in massless QED, the tree-level plus one-loop
one-particle-irreducible vertex function describing scattering of
electron in an external electromagnetic field $\bar{A}$, $ \Gamma
= \Gamma^{(0)} + \Gamma^{(1)}$, is

\vskip 3mm
\begin{center}
\includegraphics[bb=31 69 272 132,width=.5\textwidth]{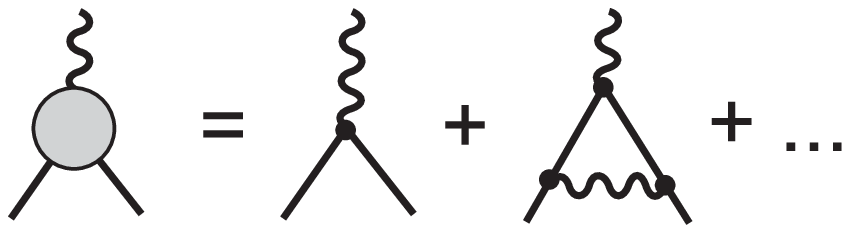}
\end{center}

\vskip 3mm If we calculate the RG operator in one-loop order
\begin{equation} \label{rgo1}
{{\cal D}}  = \frac{\partial}{\partial \log \mu} + \beta_e^{(1)}
\frac{\partial}{\partial e} - 2 \gamma_\psi^{(1)} - \gamma_A^{(1)}
,
\end{equation}
where $\beta_e^{(1)}$, $ \gamma_A^{(1)}, \gamma_\psi^{(1)}$ are
the beta-function and the anomalous dimensions  of electromagnetic
and electron fields, respectively, and apply it to $\Gamma$, we
obtain
\begin{equation} \label{rge1}
- \frac{\partial}{\partial\log\mu}\Gamma^{(1)} = \left(
\beta_e^{(1)} \frac{\partial}{\partial e} - 2 \gamma_\psi^{(1)} -
\gamma_A^{(1)}\right)~\Gamma^{(0)} + O(e^5).
\end{equation}
 Then, accounting for the
values of
\begin{equation} \label{parameter}
\beta_e^{(1)} = \frac{e^3}{12 \pi^2},\quad \gamma_A^{(1)} =
\frac{e^2}{12 \pi^2},\quad \gamma_{\psi}^{(1)} = \frac{e^2}{16
\pi^2}
\end{equation}
and the factor $e$ in  $\Gamma^{(0)}$, we observe that the first
and the last terms in the r.h.s. cancel. Since $\mu$-dependent
term  in $\Gamma^{(1)} $ is $\Gamma^{(1)}_\mu = \frac{e^3 }{16
\pi^2}\log \mu^2$, we see that Eq.(\ref{rge1}) is identity in the
order $O(e^3)$.

Next important point is that in a theory with different mass
scales the decoupling of heavy-loop contributions at the threshold
of heavy masses, $\Lambda$, results in the following property: the
running of all functions is regulated by the loops of light
particles. Therefore, the $\beta$ and $\gamma$ functions at low
energies are determined by the SM particles, only. This fact is
the consequence of the decoupling theorem
\cite{appelquist75,collins78}.

The decoupling results in the redefinition of  parameters at the
scale $\Lambda$ and removing heavy-particle loop contributions
from RG equation \cite{bando93,EPJC2000}:
\begin{eqnarray} \label{redefinition}
\lambda_a &=& \hat{\lambda}_a + a_{\hat{\lambda}_a}\log
\frac{\hat{\Lambda}}{\mu^2} + b_{\hat{\lambda}_a} \log^2
\frac{\Lambda^2}{\mu^2}  + \cdots,
\\ \nonumber
X &=& \hat{X}\left( 1 + a_{\hat{\lambda}_a}\log
\frac{\hat{\Lambda}}{\mu^2} + b_{\hat{\lambda}_a} \log^2
\frac{\Lambda^2}{\mu^2}  + \cdots\right),
\end{eqnarray}
where $\lambda_a$ and $X$ denote the parameters of the SM. They
are calculated assuming that no heavy particles are excited inside
loops.

The matching between both sets of parameters {$\lambda_a$, $X$}
and {$\hat{\lambda}_a$, $\hat{X}$} is chosen at the normalization
point $\mu \sim \Lambda$,
\begin{equation} \label{match}
\lambda_a |_{\mu = \Lambda} = \hat{\lambda}_a |_{\mu =
\Lambda},\quad X_a |_{\mu = \Lambda} = \hat{X}_a |_{\mu = \Lambda}
.
\end{equation}

The differential operator ${\cal D}$ in the RG equation is in fact
unique;  the apparently different ${\cal D}$ in both theories are
the same!

Note that if a theory with different mass scales is specified one
can freely replace the parameters {$\lambda_a$, $X$} and
{$\hat{\lambda}_a$, $\hat{X}$} by each other
\cite{bando93,EPJC2000}.

An example of the derivation and the main features of the RG
relations are shown in the Appendix for a simple model with
different mass scales.

If underlying theory is not specified, the set of
{$\hat{\lambda}_a$, $\hat{X}$} is unknown. The low energy theory
consists of the SM plus the effective Lagrangian generated by the
interactions of light particles with virtual heavy particle
states.  The low energy parameters $\lambda'_{l.}$ of these
interactions are arbitrary numbers which must be constrained by
experiments. By calculating the RG operator ${\cal D}$ and the
scattering amplitudes of light particles in this `external field'
in a chosen order of loop expansion, it is possible to obtain the
model-independent correlations between $\lambda'_{l.}$. These are
just the RE relations.

\section{The RG relations for $Z'$ boson couplings}

Let us derive the correlations between $ \tilde{Y}_{\phi_{i,1}}$,
$ \tilde{Y}_{\phi_{i,2}}, \tilde{Y}_{L,f}, \tilde{Y}_{R,f}$
appearing due to renormalizability of the underlying theory
containing $Z'$.

In our case, the RG invariance of the vertex leads to the equation
\begin{equation} \label{matchA}
{\cal D}\left(\bar{f} \Gamma_{fZ'} f \frac{1}{m_Z'}\right) = 0,
\end{equation}
where
\begin{equation} \label{rgo1A}
{{\cal D}} = \frac{d}{d \log \mu} = \frac{\partial}{\partial \log
\mu} + \sum\limits_a \beta_a \frac{\partial}{\partial \lambda}_a -
\sum\limits_{X} \gamma_X \frac{\partial}{\partial \log X},
\end{equation}
and
\begin{equation} \label{betaX}
\beta_{a} = \frac{d \lambda}{d \log \mu}, ~~\gamma_X = - \frac{d
\log X}{d \log \mu}
\end{equation}
are computed with taking into account the loops of light
particles.

Now  we derive the RG relations following from the one-loop
consideration. In accordance with the previous sections, the
one-loop  RG equation for the vertex function reads
\begin{equation} \label{rgez'}
\bar{f}\frac{\partial \Gamma^{(1)}_{f Z'} }{\partial \log \mu} f
\frac{1}{m_{Z'}} + {\cal D}^{(1)} \left(\bar{f} \Gamma^{(0)}_{f
Z'} f \frac{1}{m_{Z'}}\right) = 0,
\end{equation}
where $\Gamma^{(0)}_{f Z'}$ and $\Gamma^{(1)}_{f Z'}$ are the
tree-level and one-loop contributions to the fermion-$Z'$ vertex.
${\cal D}^{(1)} = \sum\limits_a \beta^{(1)}_a
\frac{\partial}{\partial \lambda}_a - \sum\limits_{X}
\gamma^{(1)}_X \frac{\partial}{\partial \log X}$ is the one-loop
level part of the RG operator.

\begin{figure}
\centering
\includegraphics[bb=10 24 577 705, width=.6\textwidth]{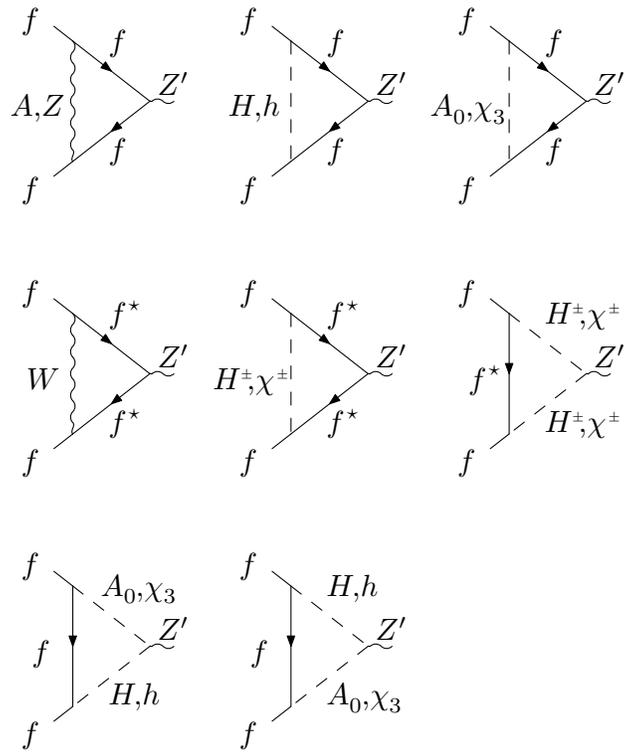}
\caption{The diagrams contributing to the divergent parts of the
$Z'ff$ vertex at the one-loop level.}\label{f-2}
\end{figure}
\begin{figure}
\centering
\includegraphics[bb=10 38 581 470, width=.6\textwidth]{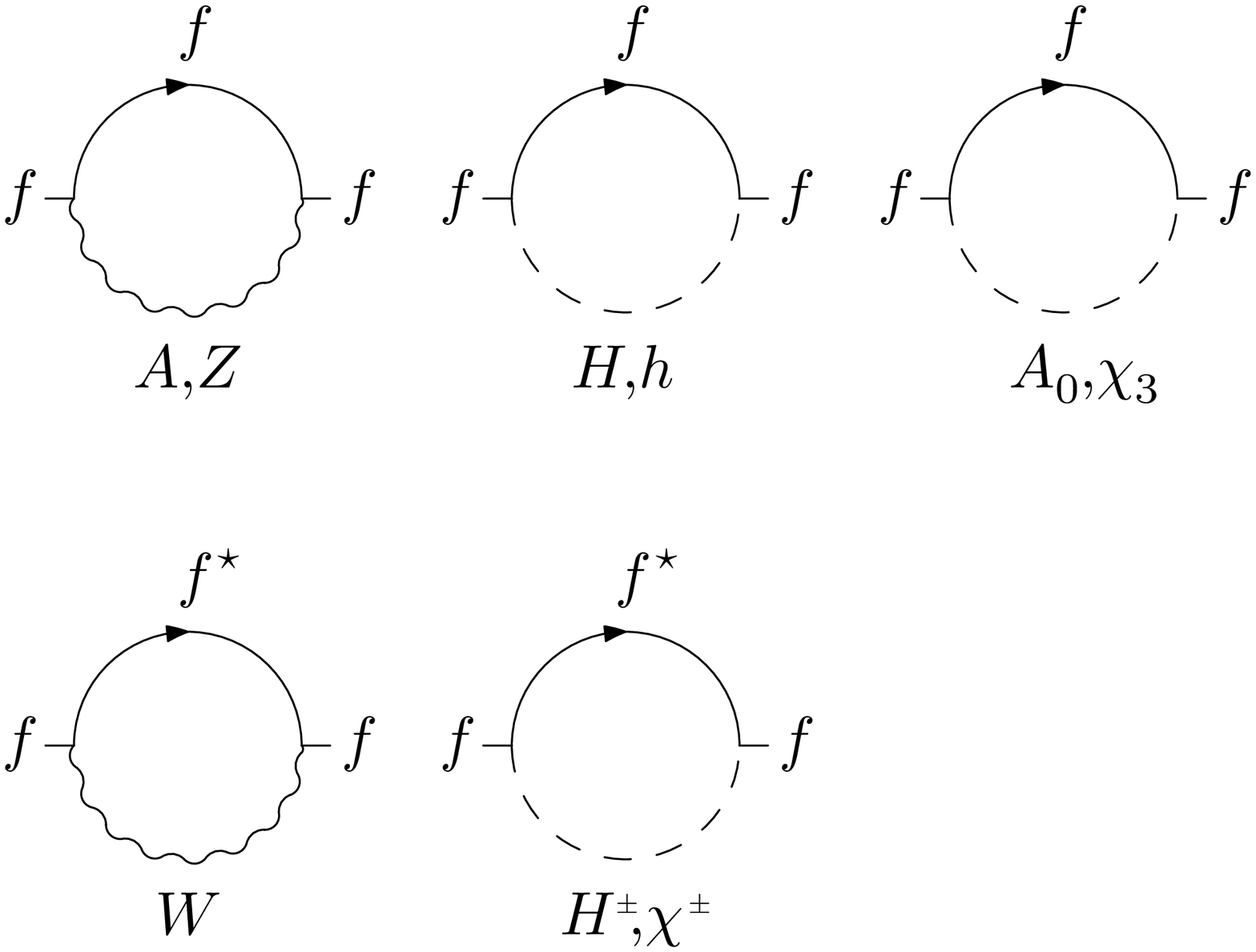}
\caption{The diagrams contributing to the fermion anomalous
dimension at the one-loop level.}\label{f-3}
\end{figure}

To calculate these functions, only the divergent parts of the
one-loop vertices are to be calculated. The corresponding diagrams
are shown in Fig. \ref{f-2}. The fermion anomalous dimensions
$\gamma^{(1)}_X$ can be calculated by using the diagrams in Fig.
\ref{f-3}. Then, Eq.(\ref{rgez'}) leads to algebraic  equations
for the parameters $ \tilde{Y}_{\phi_{i,1}}$, $
\tilde{Y}_{\phi_{i,2}}, \tilde{Y}_{L,f},$ and $\tilde{Y}_{R,f}$
which have two sets of solutions \cite{EPJC2000}:
\begin{eqnarray} \label{rgr1}
\tilde{Y}_{\phi_{2,1}} &=& \tilde{Y}_{\phi_{1,1}} = -
\tilde{Y}_{\phi_{,2}} \equiv - \tilde{Y}_{\phi_{}}, \\ \nonumber
\tilde{Y}_{L,f}&+& \tilde{Y}_{L,f^*} = 0,~ \tilde{Y}_{R,f} = 0,
\end{eqnarray}
and
\begin{eqnarray} \label{rgr2}
\tilde{Y}_{\phi_{1,1}} &=& \tilde{Y}_{\phi_{2,1}} =
\tilde{Y}_{\phi_{,2}} \equiv  \tilde{Y}_{\phi_{}}, \\ \nonumber
\tilde{Y}_{L,f}&=& \tilde{Y}_{L,f^*},~~ \tilde{Y}_{R,f} =
\tilde{Y}_{L,f} + 2 T_{3f}~ \tilde{Y}_{\phi_{}}.
\end{eqnarray}
Here $f$ and $f^*$ are the partners of the $SU(2)_L$ fermion
doublet ($l^* = \nu_l, \nu^* = l, q^*_u = q_d$ and $q^*_d = q_u$),
$T_{3f}$ is the third component of weak isospin.

The first of these relations describes the $Z'$ boson analogous to
the third component of the $SU(2)_L$ gauge field. The couplings to
the right-handed singlet are absent.

The second relation corresponds to the Abelian $Z'$. In this case
the SM Lagrangian appears to be invariant with respect to the
$\tilde{U}(1)$ group associated with the $Z'$. The last relation
in Eq.(\ref{rgr2}) ensures  the $L_\mathrm{Yuk.}$
Eq.(\ref{Lyukawa}) is to be invariant with respect to the
$\tilde{U}(1)$ transformations.

Introducing the $Z'$ couplings to the vector and axial-vector
fermion currents (\ref{av}), the last line in Eq. (\ref{rgr2})
yields
\begin{equation} \label{grgav}
v_f - a_f= v_{f^*} - a_{f^*}, \qquad a_f = T_{3f}
\tilde{g}\tilde{Y}_\phi.
\end{equation}
The couplings of the Abelian $Z'$ to the axial-vector fermion
current have  a universal absolute value proportional to the $Z'$
coupling to the scalar doublet.

These relations are model independent. In particular, they hold in
all the known models containing the Abelian $Z'$. The most
discussed models are derived from the ${\rm E}_6$ group (the so
called LR, $\chi$-$\psi$ models). The tree-level $Z'$ couplings to
the SM fermions in the models are shown in Table 1.

\begin{table}
\begin{center}
\label{zpcoup} \caption{The $Z^\prime$ couplings to the SM
fermions in the most discussed ${\rm E}_6$-based models.}
\begin{tabular}{|l|c|c|c|c|}
\hline $f$ & \multicolumn{2}{|c|}{$\chi$-$\psi$} &
\multicolumn{2}{|c|}{LR}\\
 \cline{2-5}
 & $a_f/\tilde{g}$ & $v_f/\tilde{g}$
      & $a_f/\tilde{g}$ & $v_f/\tilde{g}$
\\ \hline\hline
 &&&&\\
 $\nu$ &
 $-3\frac{\cos{\beta}}{\sqrt{40}}
  -\frac{\sin{\beta}}{\sqrt{24}}$ &
 $3\frac{\cos{\beta}}{\sqrt{40}}
  +\frac{\sin{\beta}}{\sqrt{24}}$ &
 $-\frac{1}{2\alpha}$ & $\frac{1}{2\alpha}$ \\
 &&&&\\
 $e$  &
 $-\frac{\cos{\beta}}{\sqrt{10}}
  -\frac{\sin{\beta}}{\sqrt{6}}$ &
 $2\frac{\cos{\beta}}{\sqrt{10}}$ &
 $-\frac{\alpha}{2}$ & $\frac{1}{\alpha}-\frac{\alpha}{2}$ \\
 &&&&\\
 $q_u$ &
 $\frac{\cos{\beta}}{\sqrt{10}}
 -\frac{\sin{\beta}}{\sqrt{6}}$ & 0 &
 $\frac{\alpha}{2}$&$-\frac{1}{3\alpha}+\frac{\alpha}{2}$ \\
 &&&&\\
 $q_d$ &
 $-\frac{\cos{\beta}}{\sqrt{10}}
  -\frac{\sin{\beta}}{\sqrt{6}}$ &
 $-2\frac{\cos{\beta}}{\sqrt{10}}$ &
 $-\frac{\alpha}{2}$ & $-\frac{1}{3\alpha}-\frac{\alpha}{2}$
\\ &&&&\\
 \hline
\end{tabular}
\end{center}
\end{table}

The ${\rm E}_6$-symmetry breaking scheme
\[
{\rm E}_6\to{\rm SO}(10)\times{\rm U}(1)_\psi,\quad {\rm
SO}(10)\to{\rm SU}(3)_c\times{\rm SU}(2)_L \times{\rm
SU}(2)_R\times{\rm U}(1)_{B-L}.
\]
leads to the so called left-right (LR) model. Another scheme,
\[
{\rm E}_6\to{\rm SO}(10)\times{\rm U}(1)_\psi\to{\rm SU}(5)\times
{\rm U}(1)_\chi\times{\rm U}(1)_\psi,
\]
predicts the Abelian $Z^\prime$, which is a linear combination of
the neutral vector bosons $\psi$ and $\chi$,
\[
Z^\prime=\chi\cos{\beta} +\psi\sin{\beta}
\]
with the mixing angle $\beta$. If we suppose only one $Z'$ boson
at low energies, the $\psi$ boson should be much heavier than the
$\chi$ field. In this case, the field $\psi$ is decoupled and
$\beta\to 0$. As it is seen, both the LR and the $\chi$-$\psi$
models (with $\beta=0$ to avoid two $Z'$ bosons with the same
scale of masses) satisfy the RG relations (\ref{rgr2}) except for
neutrinos. Let us explain this discrepancy. It is usually supposed
in theories based on the ${\rm E}_6$ group that the Yukawa terms
responsible for generation of the Dirac masses of neutrinos must
be set to zero \cite{Hewett}. Therefore, the terms proportional to
the Yukawa couplings vanish in the renormalization group equation,
and there are no RG relations for the $Z^\prime$ interactions with
the neutrino axial-vector currents. In this case the couplings
$a_\nu$ given in Table 1
are not restricted by the RG relations.

\section{Implication of the RG relations}

LEP collaborations applied model dependent search for $Z'$ and
have obtained low bounds on the mass $ m_{Z'} \geq 400 - 800$ GeV
dependently on a specific model \cite{EWWG,OPAL,DELPHI}.

In our analysis, we consider the SM with the additional effective
$Z'$ interactions (\ref{Lf}), (\ref{Lscal}), (\ref{Lyukawa}) as a
low energy theory. The parameters $a_f, v_f$ and $m_{Z'}$ must be
fitted in experiments. The RG relations give a possibility:

\begin{enumerate}
    \item reduce the number of fitted parameters;
    \item determine kinematics of the processes;
    \item introduce observables which uniquely pick out the $Z'$ signals.
\end{enumerate}

The RG relations (\ref{rgr2}) influences the $Z - Z'$ mixing
(\ref{2}). The axial-vector coupling determines also the coupling
to the scalar doublet and, consequently, the mixing angle. As a
result, the number of independent couplings is significantly
reduced.

In what follows, both types of the RG relations (\ref{rgr1}) and
(\ref{rgr2}) will be used  in order to search for signals of the
$Z'$ gauge boson.

\section{$Z'$ search in $e^+e^-\to\mu^+\mu^-,\tau^+\tau^-$ processes}

\subsection{The differential cross section}

Let us consider the processes $e^+e^-\to l^+l^-$ ($l=\mu,\tau$)
with the non-polarized initial- and final-state fermions. In order
to introduce the observable which selects the signal of the
Abelian $Z'$ boson we need to compute the differential
cross-sections of the processes up to the one-loop level.

The lower-order diagrams describe the neutral vector boson
exchange in the $s$-channel ($e^+e^-\to V^\ast\to l^+l^-$,
$V=A,Z,Z'$). As for the one-loop corrections, two classes of
diagrams are taken into account. The first one includes the pure
SM graphs (the mass operators, the vertex corrections, and the
boxes). The second set of the one-loop diagrams improves the
Born-level $Z'$-exchange amplitude by ``dressing'' the $Z'$
propagator and and the $Z'$--fermion vertices. We assume that $Z'$
states are not excited inside loops. Such an approximation means
that the $Z'$-boson is completely decoupled. Then, the
differential cross-section consists of the squared tree-level
amplitude and the term from the interference of the tree-level and
the one-loop amplitudes. To obtain an infrared-finite result, we
also take into account the processes with the soft-photon emission
in the initial and final states.

In the lower order in $m^{-2}_{Z'}$ the $Z'$ contributions to the
differential cross-section of the process $e^+e^-\to l^+l^-$ are
expressed in terms of four-fermion contact couplings, only. If one
takes into consideration the higher-order corrections in
$m^{-2}_{Z'}$, it becomes possible to estimate separately the
$Z'$-induced contact couplings and the $Z'$ mass \cite{rizzo}. In
the present analysis we keep the terms of order $O(m^{-4}_{Z'})$
to fit both of these parameters.

Expanding the differential cross-section in the inverse $Z'$ mass
and neglecting the terms of order $O(m^{-6}_{Z'})$, we have
\begin{eqnarray}
\frac{d\sigma_l(s)}{dz} &=& \frac{d\sigma_l^{\rm SM}(s)}{dz}
+\sum_{i=1}^{7}\sum_{j=1}^{i}
\left[A_{ij}^l(s,z)+B_{ij}^l(s,z)\zeta\right]x_{i}x_{j}
\nonumber\\&&
+\sum_{i=1}^{7}\sum_{j=1}^{i}\sum_{k=1}^{j}\sum_{n=1}^{k}
C_{ijkn}^l(s,z)x_{i}x_{j}x_{k}x_{n},
\end{eqnarray}
where the dimensionless quantities
\begin{eqnarray}
 \zeta = \frac{m^2_Z}{m^2_{Z'}},\quad
 (x_1,x_2,x_3,x_4,x_5,x_6,x_7) =
 (\bar{a},\bar{v}_e,\bar{v}_\mu,\bar{v}_\tau,\bar{v}_d,\bar{v}_s,\bar{v}_b)
\end{eqnarray}
are introduced. Since the axial-vector couplings of the Abelian
$Z'$ boson are universal, we use the shorthand notation
$\bar{a}=\bar{a}_e$. In what follows the index $l=\mu,\tau$
denotes the final-state lepton.

The coefficients $A$, $B$, $C$ are determined by the SM couplings
and masses. Each factor may include the tree-level contribution,
the one-loop correction and the term describing the soft-photon
emission. The factors $A$ describe the leading-order contribution,
whereas others correspond to the higher order corrections in
$m^{-2}_{Z'}$.

\subsection{The observable}

To take into consideration the correlations (\ref{2}) we introduce
the observable $\sigma_l(z)$ defined as the difference of cross
sections integrated in some ranges of the scattering angle
$\theta$ \cite{PRD2000,YAF2004}:
\begin{eqnarray}\label{eq8}
 \sigma_l(z)
 &\equiv&\int\nolimits_z^1
  \frac{d\sigma_l}{d\cos\theta}d\cos\theta
 -\int\nolimits_{-1}^z
  \frac{d\sigma_l}{d\cos\theta}d\cos\theta,
\end{eqnarray}
where $z$ stands for the cosine of the boundary angle. The idea of
introducing the $z$-dependent observable (\ref{eq8}) is to choose
the value of the kinematic parameter $z$ in such a way that to
pick up the characteristic features of the Abelian $Z'$ signals.

The deviation of the observable from its SM value can be derived
by the angular integration of the differential cross-section and
has the form:
\begin{eqnarray}
\Delta\sigma_l(z) &=& \sigma_l(z) - \sigma^{\rm SM}_l(z)
=\sum_{i=1}^{7}\sum_{j=1}^{i}
\left[\tilde{A}_{ij}^l(s,z)+\tilde{B}_{ij}^l(s,z)\zeta\right]x_{i}x_{j}
\nonumber\\&&
+\sum_{i=1}^{7}\sum_{j=1}^{i}\sum_{k=1}^{j}\sum_{n=1}^{k}
\tilde{C}_{ijkn}^l(s,z)x_{i}x_{j}x_{k}x_{n}.
\end{eqnarray}

\begin{figure}
\begin{center}
  \includegraphics[bb= 0 0 600 600, width=.5\textwidth]{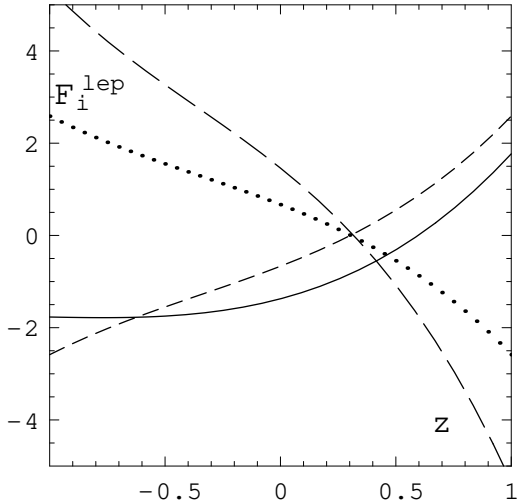}
 \caption{The leptonic functions
  $F^{l}_{0}$ (the solid curve),
  $F^{l}_{1}$ (the long-dashed curve),
  $F^{l}_{2}$ (the dashed curve), and
  $F^{l}_{3}$ (the dotted curve) at $\sqrt{s}=500$ GeV.}
 \label{fig:2f}
 \end{center}
\end{figure}

Then let us introduce the quantity
$\Delta\sigma\left(z\right)\equiv \sigma\left(z\right)
-{\sigma}_{SM}\left(z\right)$, which owing to the relations
(\ref{rgr2}) can be written in the form
\begin{eqnarray}\label{obs:6}
 \Delta\sigma_f(z)
 &=&\frac{\alpha N_f}{8}\frac{g^2_{Z^\prime}}{m^2_{Z^\prime}}
  \left[
  F^f_0(z,s) \tilde{Y}^2_\phi
  +2 F^f_1(z,s) T_{3f}\tilde{Y}_{L,f}\tilde{Y}_{L,e}  \right.
 \nonumber\\
  &&
  +\left.
  2 F^f_2(z,s) T_{3f}\tilde{Y}_{L,f}\tilde{Y}_\phi
  + F^f_3(z,s) \tilde{Y}_{L,e}\tilde{Y}_\phi \right].
\end{eqnarray}
The factor functions $F^f_i(z,s)$ depend on the fermion type
through the $|Q_f|$, only. In Fig. \ref{fig:2f} they are shown as
the functions of $z$ for $\sqrt{s}=500$ GeV. The leading
contributions to $F^f_i(z,s)$,
\begin{eqnarray}\label{obs:7}
 F^f_0(z,s)&=&
  -\frac{4}{3}\left|Q_f\right|
  \left(1 -z -z^{2} -\frac{z^{3}}{3}\right)
+O\left(\frac{m^2_Z}{s}\right),
 \nonumber\\
 F^f_1(z,s)&=&\frac{4}{3}
   \left[1 -z^{2} -\left|Q_f\right|
   \left(3z+z^{3}\right)\right]
+O\left(\frac{m^2_Z}{s}\right),
 \nonumber\\
 F^f_2(z,s)&=&
  -\frac{2}{3}\left(1-z^{2}\right)
   +\frac{2}{9}\left(3z+z^{3}\right)
   \left(4\left|Q_f\right|-1\right)
      +O\left(\frac{m^2_Z}{s}\right),
 \nonumber\\
 F^f_3(z,s)&=&\frac{2}{3}\left|Q_f\right|
   \left(1-3z-z^{2}-z^{3}\right)
+O\left(\frac{m^2_Z}{s}\right),
\end{eqnarray}
are given by the $Z^\prime$ exchange diagram $e^-e^+\to
Z'\to\bar{f}f$, since the $Z$-$Z'$ mixing contribution to the $Z$
exchange diagram is suppressed by the factor $m^2_Z/s$.

From Eqs. (\ref{obs:7}) one can see that the leading contributions
to the leptonic factors $F^l_1$, $F^l_2$, $F^l_3$ are found to be
proportional to the same polynomial in $z$. This is the
characteristic feature of the leptonic functions $F^l_i$
originating due to the kinematic properties of fermionic currents
and the specific values of the SM leptonic charges. Therefore, it
is possible to choose the value of $z=z^\ast$ which switches off
three leptonic factors $F^l_1$, $F^l_2$, $F^l_3$ simultaneously.
Moreover, the quark function $F^q_3$ in the lower order is
proportional to the leptonic one and therefore is switched off,
too. As is seen from Fig. \ref{fig:2f}, the appropriate value of
$z^\ast$ is about $\sim 0.3$. By choosing this value of $z^\ast$
one can simplify Eq. (\ref{obs:6}). It is also follows from Eq.
(\ref{obs:6}) that neglecting the factors $F^l_1$, $F^l_2$,
$F^l_3$ one obtains the sign definite quantity $\Delta \sigma_l
(z^\ast)\sim \tilde{Y}^2_\phi\sim\bar{a}^2$.

There is an interval of values of the boundary angle, at which the
factors $\tilde{A}^l_{11}$, $\tilde{B}^l_{11}$, and
$\tilde{C}^l_{1111}$ at the sign-definite parameters $\bar{a}^2$,
$\bar{a}^2\zeta$, and $\bar{a}^4$ contribute more than 95\% of the
observable value. It gives a possibility to construct the
sign-definite observable $\Delta\sigma_l(z^*)<0$ by specifying the
proper value of $z^*$.

In general, one could choose the boundary angle $z^*$ in different
schemes. If just a few number of tree-level four-fermion contact
couplings are considered, one can specify $z^*$ in order to cancel
the factor at the vector-vector coupling. However, if one-loop
corrections are taken into account there are a large number of
additional contact couplings. So, we have to define some
quantitative criterion $F(z)$ to estimate the contributions from
sign-definite factors at a given value of the boundary angle $z$.
Maximizing the criterion, one could derive the value $z^*$, which
corresponds to the sign-definite observable $\Delta\sigma_l(z^*)$.
Since the observable is linear in the coefficients $A$, $B$, and
$C$, we introduce the following criterion,
\begin{equation}
F=\frac{|\tilde{A}_{11}|+\omega_B |\tilde{B}_{11}| + \omega_C
|\tilde{C}_{1111}|}
 {\sum\limits_{{\rm all}~\tilde{A}}\left|\tilde{A}_{ij}\right|
  +\omega_B\sum\limits_{{\rm all}~\tilde{B}}\left|\tilde{B}_{ij}\right|
  +\omega_C\sum\limits_{{\rm all}~\tilde{C}}\left|\tilde{C}_{ijkn}\right|},
\end{equation}
where the positive `weights' $\omega_B\sim\zeta$ and
$\omega_C\sim\epsilon$ take into account the order of each term in
the inverse $Z'$ mass.

The numeric values of the `weights' $\omega_B$ and $\omega_C$ can
be taken from the present day bounds on the contact couplings
\cite{EWWG}. As the computation shows, the value of $z^*$ with the
accuracy $10^{-3}$ depends on the order of the `weight'
magnitudes, only. So, in what follows we take $\omega_B\sim
4\times 10^{-3}$ and $\omega_C\sim 4\times 10^{-5}$.

The function $z^\ast(s)$ is the decreasing function of the
center-of-mass energy. It is tabulated for the LEP2 energies in
Table \ref{tobsmu}. The corresponding values of the maximized
function $F$ are within the interval $0.95<F<0.96$.

\begin{table}
\caption{The boundary angle $z^*$ and the coefficients in the
observable $\Delta\sigma_l(z^*)$ for the scattering into $\mu$ and
$\tau$ pairs at the one-loop level. }\label{tobsmu}\centering
\begin{center}
\begin{tabular}{|c|c|c|c|c|c|c|c|c|}\hline
  &  \multicolumn{4}{|c|}{$\mu^+\mu^-$} &  \multicolumn{4}{|c|}{$\tau^+\tau^-$} \\ \cline{2-9}
 $\sqrt{s}$, GeV & $z^*$ & $\tilde{A}_{11}$ & $\tilde{B}_{11}$ & $\tilde{C}_{1111}$
 & $z^*$ & $\tilde{A}_{11}$ & $\tilde{B}_{11}$ & $\tilde{C}_{1111}$ \\\hline\hline
 130 & 0.450 & -729 & -1792 & -19636 & 0.460 & -687 & -1664 & -25782\\
 136 & 0.439 & -709 & -1859 & -16880 & 0.442 & -688 & -1779 & -20784\\
 161 & 0.400 & -643 & -2183 & -6890 & 0.400 & -625 & -2097 & -10993\\
 172 & 0.390 & -619 & -4099 & -4099 & 0.391 & -601 & -2263 & -8382\\
 183 & 0.383 & -599 & -2545 & -1334 & 0.385 & -571 & -2402 & -7580\\
 189 & 0.380 & -586 & -2635 & -495 & 0.380 & -568 & -2533 & -5135\\
 192 & 0.380 & -579 & -2681 & -63 & 0.380 & -562 & -2578 & -4769\\
 196 & 0.380 & -571 & -2745 & -528 & 0.379 & -554 & -2640 & -4272\\
 200 & 0.378 & -564 & -2811 & -1137 & 0.378 & -547 & -2704 & -3761\\
 202 & 0.376 & -560 & -2845 & -1448 & 0.377 & -543 & -2736 & -3501\\
 205 & 0.374 & -555 & -2897 & -1923 & 0.374 & -548 & -2834 & -1292\\
 207 & 0.372 & -552 & -2932 & -2245 & 0.372 & -544 & -2868 & -1010\\
 \hline
\end{tabular}
\end{center}
\end{table}

Since $\tilde{A}^l_{11}(s,z^*)<0$, $\tilde{B}^l_{11}(s,z^*)<0$ and
$\tilde{C}^l_{1111}(s,z^*)<0$, the observable
\begin{equation}\label{7dfg}
 \Delta\sigma_l(z^*)=
 \left[\tilde{A}^l_{11}(s,z^*) +\zeta\tilde{B}^l_{11}(s,z^*)
 \right]\bar{a}^2 + \tilde{C}^l_{1111}(s,z^*)\bar{a}^4
\end{equation}
is negative with the accuracy 4--5\%. Since this property follows
from the RG relations (\ref{2}) for the Abelian $Z'$ boson, the
observable $\Delta\sigma_l(z^*)$ selects the model-indepen\-dent
signal of this particle in the processes $e^+e^-\to l^+l^-$. It
allows to use the data on scattering into $\mu\mu$ and $\tau\tau$
pairs in order to estimate the Abelian $Z'$ coupling to the
axial-vector lepton currents.

Although the observable can be computed from the differential
cross-sections directly, it is also possible to recalculate it
from the total cross-sections and the forward-backward
asymmetries. The recalculation procedure has the proper
theoretical accuracy. Nevertheless, it allows to reduce the
experimental errors on the observable, since the published data on
the total cross-sections and the forward-backward asymmetries are
more precise than the data on the differential cross-sections.

The recalculation is based on the fact that the differential
cross-section can be approximated with a good accuracy by the
two-parametric polynomial in the cosine of the scattering angle
$z$:
\begin{equation}
\frac{d\sigma_l(s)}{dz} = \frac{d\sigma_l^{\rm SM}(s)}{dz} +
(1+z^2)\beta_l + z \eta_l +\delta_l(z),
\end{equation}
where $\delta_l(z)$ measures the difference between the exact and
the approximated cross-sections. The approximated cross-section
reproduces the exact one in the limit of the massless initial- and
final-state leptons and if one neglects the contributions of the
box diagrams.

Performing the angular integration, it is easy to obtain the
expression for the observable:
\begin{equation}
\Delta\sigma_l(z^*)=\sigma_l(z^*) - \sigma_l^{\rm SM}(z^*) =
(1-z^{*2})\eta_l -\frac{2\beta_l}{9} z^*(3+z^{*2})
+\tilde\delta_l(z^*),
\end{equation}
and for the total and the forward-backward cross-sections:
\begin{eqnarray}
\Delta\sigma^{\rm T}_l &=&\sigma^{\rm T}_l- \sigma_l^{\rm T,SM}
=\frac{8\beta_l}{9}
 +\tilde\delta_l(-1), \nonumber\\ \Delta\sigma^{\rm FB}_l &=&
\sigma^{\rm FB}_l-\sigma_l^{\rm FB,SM} =\eta_l +\tilde\delta_l(0).
\end{eqnarray}
Then, the factors $\beta_l$ and $\eta_l$ can be eliminated from
the observable:
\begin{eqnarray}
\Delta\sigma_l(z^*) &=& (1-z^{*2})\Delta\sigma^{\rm FB}_l
-\frac{3}{12} z^*(3+z^{*2})\Delta\sigma^{\rm T}_l +\xi_l.
\end{eqnarray}
The quantity $\xi_l$,
\begin{eqnarray}
\xi_l&=&\tilde\delta_l(z^*) -(1-z^{*2})\tilde\delta_l(0)
+\frac{3}{12} z^*(3+z^{*2})\tilde\delta_l(-1),
\end{eqnarray}
measures the theoretical accuracy of the approximation.

The forward-backward cross-section is related to the total one and
the forward-backward asymmetry by means of the following
expression
\begin{eqnarray}
\Delta\sigma^{\rm FB}_l &=& \Delta\sigma^{\rm T}_l \,\, A^{\rm
FB}_l +\sigma_l^{\rm T,SM} \,\, \Delta A^{\rm FB}_l.
\end{eqnarray}
As the computation shows, $\tilde\delta_l(z^*)\simeq 0.01
\Delta\sigma_{l}(z^*)$, $\tilde\delta_l(0)\simeq 0.007
\Delta\sigma^{\rm FB}_l$, and $\tilde\delta_l(-1)\simeq -0.07
\Delta\sigma^{\rm T}_l$ at the LEP2 energies. Taking into account
the experimental values of the total cross-sections and the
forward-backward asymmetries at the LEP2 energies
($\Delta\sigma^{\rm T}_l\simeq 0.1$pb, $\sigma_l^{\rm T,SM}\simeq
2.7$pb, $\Delta A_l^{\rm FB}\simeq 0.04$, $A_l^{\rm FB}\simeq
0.5$), one can estimate the theoretical error as $\xi_l\simeq
0.003{\rm pb}$. At the same time, the corresponding statistical
uncertainties on the observable are larger than 0.06pb. Thus, the
proposed approximation is quite good and can be successfully used
to obtain more accurate experimental values of the observable.

\subsection{Data fit}

To search for the model-independent signals of the Abelian
$Z'$-boson we will analyze the introduced observable
$\Delta\sigma_l (z^\ast)$ on the base of the LEP2 data set. In the
lower order in $m^{-2}_{Z'}$ the observable (\ref{7dfg}) depends
on one flavor-independent parameter $\bar{a}^2$,
\begin{equation}
 \Delta\sigma^{\rm th}_l(z^*)=
 \tilde{A}^l_{11}(s,z^*)\bar{a}^2 + \tilde{C}^l_{1111}(s,z^*)\bar{a}^4,
\end{equation}
which can be fitted from the experimental values of
$\Delta\sigma_\mu (z^\ast)$ and $\Delta\sigma_\tau (z^\ast)$. As
we noted above, the sign of the fitted parameter ($\bar{a}^2 >0$)
is a characteristic feature of the Abelian $Z'$ signal.

In what follows we will apply the usual fit method based on the
likelihood function. The central value of $\bar{a}^2$ is obtained
by the minimization of the $\chi^2$-function:
\begin{equation}
\chi^2(\bar{a}^2) = \sum_{n} \frac{\left[\Delta\sigma^{\rm
ex}_{\mu,n}(z^*)- \Delta\sigma^{\rm th}_\mu(z^*)\right]^2}
{\delta\sigma^{\rm ex}_{\mu,n}(z^*)^2},
\end{equation}
where the sum runs over the experimental points entering a data
set chosen. The $1\sigma$ CL interval $(b_1,b_2)$ for the fitted
parameter is derived by means of the likelihood function ${\cal
L}(\bar{a}^2)\propto\exp[-\chi^2(\bar{a}^2)/2]$. It is determined
by the equations:
\begin{equation}
\int\nolimits_{b_1}^{b_2}{\cal L}(\epsilon ')d\epsilon ' = 0.68,
 \quad
{\cal L}(b_1)={\cal L}(b_2).
\end{equation}

To compare our results with those of Refs. \cite{EWWG} we
introduce the contact interaction scale
\begin{equation}
\Lambda^2 = 4m^2_Z\bar{a}^{-2}.
\end{equation}
This normalization of contact couplings is admitted in Refs.
\cite{EWWG}. We use again the likelihood method to determine a
one-sided lower limit on the scale $\Lambda$ at the 95\% CL. It is
derived by the integration of the likelihood function over the
physically allowed region $\bar{a}^2>0$. The strict definition is
\begin{equation}
\Lambda=2m_Z (\epsilon^*)^{-1/2}, \quad \int_{0}^{\epsilon^*}{\cal
L}(\epsilon ')d\epsilon ' = 0.95\int_{0}^{\infty}{\cal L}(\epsilon
')d\epsilon '.
\end{equation}

We also introduce the probability of the Abelian $Z'$ signal as
the integral of the likelihood function over the positive values
of $\bar{a}^2$:
\begin{equation}
P=\int\nolimits_{0}^{\infty} L(\epsilon ')d\epsilon '.
\end{equation}

Actually, the fitted value of the contact coupling $\bar{a}^2$
originates mainly from the leading-order term in the inverse $Z'$
mass in Eq. (\ref{7dfg}). The analysis of the higher-order terms
allows to estimate the constraints on the $Z'$ mass alone.
Substituting $\bar{a}^2$ in the observable (\ref{7dfg}) by its
fitted central value, one obtains the expression
\begin{equation}
 \Delta\sigma_l(z^*)=
 \left[\tilde{A}^l_{11}(s,z^*)
  +\zeta \tilde{B}^l_{11}(s,z^*) \right]\bar{a}_\mathrm{fitted}^2 +
 \tilde{C}^l_{1111}(s,z^*)\bar{a}_\mathrm{fitted}^4,
\end{equation}
which depends on the parameter $\zeta=m^2_Z/m^2_{Z'}$. Then, the
central value of this parameter and the corresponding 1$\sigma$ CL
interval are derived in the same way as those for $\bar{a}^2$.

To fit the parameters $\bar{a}^2$ and $\zeta$ we start with the
LEP2 data on the total cross-sections and the forward-backward
asymmetries \cite{EWWG}. The corresponding values of the
observable $\Delta\sigma_l(z^\ast)$ with their uncertainties
$\delta\sigma_l(z^\ast)$ are calculated from the data by means of
the following relations:
\begin{eqnarray}
 \Delta\sigma_l(z^\ast)
 &=&
 \left[
 A_l^{\rm FB}\left(1-z^{\ast 2}\right)
 -\frac{z^\ast}{4}\left(3 +z^{\ast 2}\right)
 \right] \Delta\sigma_l^{\rm T}
 \nonumber\\&&
 + \left(1 - z^{\ast 2}\right)
  \sigma_l^{\rm T,SM} \Delta A_l^{\rm FB},
 \nonumber\\
 \delta\sigma_l(z^\ast)^2
 &=&
 {\left[
 A_l^{\rm FB}\left(1-z^{\ast 2}\right)
 -\frac{z^\ast}{4}\left(3 +z^{\ast 2}\right)
 \right]}^2 (\delta\sigma_l^{\rm T})^2
 \nonumber\\&&
 +{\left[
 \left(1 - z^{\ast 2}\right)
 \sigma_{l}^{\rm T,SM}
 \right]}^2 (\delta A_l^{\rm FB})^2.
\end{eqnarray}

We perform the fits assuming several data sets, including the
$\mu\mu$, $\tau\tau$, and the complete $\mu\mu$ and $\tau\tau$
data, respectively. The results are presented in Table
\ref{tfitrecalc}.
\begin{table}
\caption{The contact coupling $\bar{a}^2$ with the 68\% CL
uncertainty, the 95\% CL lower limit on the scale $\Lambda$, the
probability of the $Z'$ signal, $P$, and the value of
$\zeta=m^2_Z/m^2_{Z'}$ as a result of the fit of the observable
recalculated from the total cross-sections and the
forward-backward asymmetries. }\label{tfitrecalc}\centering
\begin{tabular}{|l|c|c|c|c|}
\hline Data set & $\bar{a}^2$ & $\Lambda$, TeV & $P$ & $\zeta$
 \\ \hline\hline
 $\mu\mu$ & $0.0000366^{+0.0000489}_{-0.0000486}$
 & 16.4
 & 0.77
 & $0.009\pm 0.278$
 \\
 $\tau\tau$ & $-0.0000266^{+0.0000643}_{-0.0000639}$
 & 17.4
 & 0.34
 & $-0.001\pm 0.501$
 \\
 $\mu\mu$ and $\tau\tau$ & $0.0000133^{+0.0000389}_{-0.0000387}$
 & 19.7
 & 0.63
 & $0.017\pm 0.609$
 \\ \hline
\end{tabular}
\end{table}
As is seen, the more precise $\mu\mu$ data demonstrate the signal
of about 1$\sigma$ level. It corresponds to the Abelian $Z'$-boson
with the mass of order 1.2--1.5 TeV if one assumes the value of
$\tilde\alpha=\tilde{g}^2/4\pi$ to be in the interval 0.01--0.02.
No signal is found by the analysis of the $\tau\tau$
cross-sections. The combined fit of the $\mu\mu$ and $\tau\tau$
data leads to the signal below the 1$\sigma$ CL.

Being governed by the next-to-leading contributions in
$m^{-2}_{Z'}$, the fitted values of $\zeta$ are characterized by
significant errors. The $\mu\mu$ data set gives the central value
which corresponds to $m_{Z'}\simeq 1.1$ TeV.

We also perform a separate fit of the parameters based on the
direct calculation of the observable from the differential
cross-sections. The experimental uncertainties of the data on the
differential cross-sections are of one order larger than the
corresponding errors of the total cross-sections and the
forward-backward asymmetries. These data also provide the larger
values of the contact coupling $\bar{a}^2$. As for the more
precise $\mu\mu$ data, three of the LEP2 Collaborations
demonstrate positive values of $\bar{a}^2$. The combined
$\bar{a}^2$ is also positive and remains practically unchanged by
the incorporation of the $\tau\tau$ data.

As it was mentioned in the previous section, the indirect
computation of the observable from the total cross-sections and
the forward-backward asymmetries inspires some insufficient
theoretical uncertainty about 2\% of the statistical one. It also
increases the statistical error because of the recalculation
procedure. Nevertheless, the uncertainty of the fitted parameter
$\bar{a}^2$ within the recalculation scheme is of one order less
than that for the direct computation from the differential
cross-sections. This difference is explained by the different
accuracy of the available experimental data on the differential
and the total cross-sections.

\section{Search for $Z'$ in $e^+e^-\to e^+e^-$ process}

\subsection{The differential cross-section}

In our analysis, as the SM values of the cross-sections we use the
quantities calculated by the LEP2 collaborations
\cite{OPAL,DELPHI,ALEPH,L3}. They account for either the one-loop
radiative corrections or initial and final state radiation effects
(together with the event selection rules, which are specific for
each experiment). As it is reported by the DELPHI Collaboration,
there is a theoretical error of the SM values of about 2\%. In our
analysis this error is added to the statistical and systematic
ones for all the Collaborations. As it was checked, the fit
results are practically insensitive to accounting for this error.

The deviation from the SM is computed in the improved Born
approximation. This approximation is sufficient for our analysis
leading to the systematic error of the fit results less than 5-10
per cents.

\begin{figure}
\centering
  \includegraphics[bb= 0 0 600 400, width=.4\textwidth]{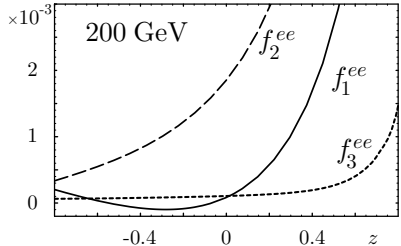}
  \caption{The factors at the $Z'$ couplings in the differential cross-section
  of the Bhabha process.}\label{fig:0}
\end{figure}

The deviation from the SM of the differential cross-section for
the process $e^+e^-\to\ell^+\ell^-$ can be expressed through
various quadratic combinations of couplings $a=a_e$, $v_e$,
$v_\mu$, $v_\tau$. For the Bhabha process it reads
\begin{equation}\label{4}
\frac{d\sigma}{dz}-\frac{d\sigma^\mathrm{SM}}{dz} =
f^{ee}_1(z)\frac{a^2}{m_{Z'}^2} +
f^{ee}_2(z)\frac{v_e^2}{m_{Z'}^2} +
f^{ee}_3(z)\frac{a v_e}{m_{Z'}^2},
\end{equation}
where the factors are known functions of the center-of-mass energy
and the cosine of the electron scattering angle $z$ plotted in
Fig. \ref{fig:0}.
The deviation of the cross-section for  $e^+e^-\to\mu^+\mu^-$
($\tau^+\tau^-$) processes has a similar form
\begin{eqnarray}\label{5}
\frac{d\sigma}{dz}-\frac{d\sigma^\mathrm{SM}}{dz} &=&
f^{\mu\mu}_1(z)\frac{a^2}{m_{Z'}^2} +
f^{\mu\mu}_2(z)\frac{v_e v_\mu}{m_{Z'}^2} +
\nonumber\\ &&+
f^{\mu\mu}_3(z)\frac{a v_e}{m_{Z'}^2} +
f^{\mu\mu}_4(z)\frac{a v_\mu}{m_{Z'}^2}.
\end{eqnarray}
Eqs. (\ref{4})--(\ref{5}) are our definition of the $Z'$ signal.

Note again that the cross-sections in Eqs. (\ref{4})--(\ref{5})
account for the relations (\ref{3}) through the functions
$f_1(z)$, $f_3(z)$, $f_4(z)$, since the coupling $\tilde{Y}_\phi$
(the mixing angle $\theta_0$) is substituted by the axial coupling
constant $a$. Usually, when a four-fermion effective Lagrangian is
applied to describe physics beyond the SM \cite{Pankov}, this
dependence on the scalar field coupling is neglected at all.
However, in our case, when we are interested in searching for
signals of the $Z'$-boson on the base of the effective low-energy
Lagrangian (\ref{Lf})--(\ref{Lyukawa}), these contributions to the
cross-section are essential.

\subsection{One-parameter fit}

The factor $f^{ee}_2(z)$ is positive monotonic function of $z$
(see Fig. 4 for the center-of-mass energies $\sqrt{s}=200$ GeV.
The same behavior is observed for higher energies). Such a
property allows one to choose $f^{ee}_2(z)$ as a normalization
factor for the differential cross section. Then the normalized
deviation of the differential cross-section reads \cite{PRD2004}
\begin{eqnarray}\label{ncs}
\frac{d\tilde\sigma}{dz}&=& \frac{m_Z^2}{{4\pi}f^{ee}_2(z)}
\Delta\,\frac{d\sigma}{dz} = \nonumber\\&&
 \bar{v}^2 +
F_{a}(\sqrt{s},z) \bar{a}^2 + F_{av}(\sqrt{s},z)\bar{a}\bar{v}
+\ldots ,
\end{eqnarray}
and the normalized factors are shown in Fig 5.
\begin{figure}
\centering
\includegraphics[bb= 0 0 500 350 ,width=70mm]{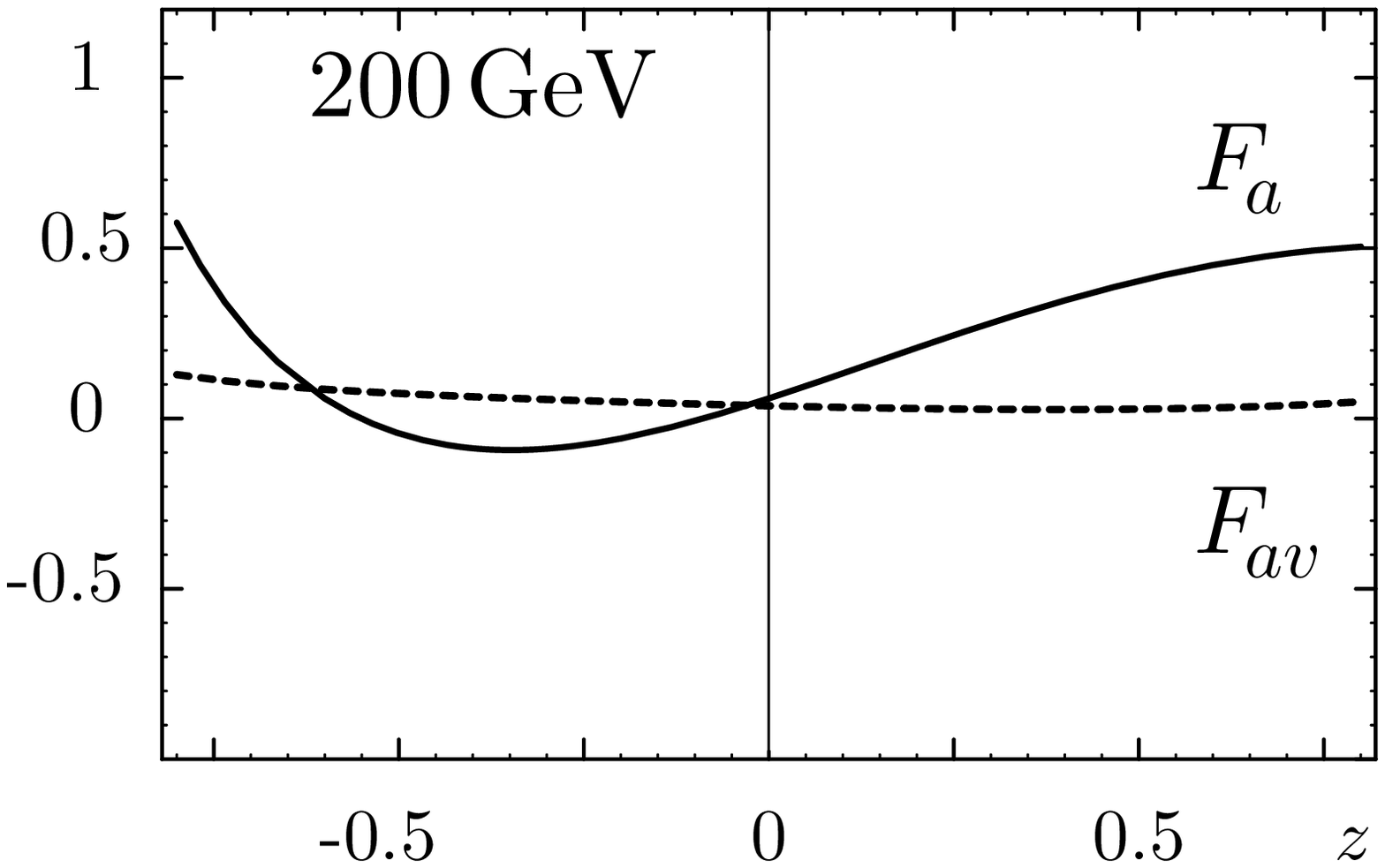}
\\
\includegraphics[bb= 0 0 500 350 ,width=70mm]{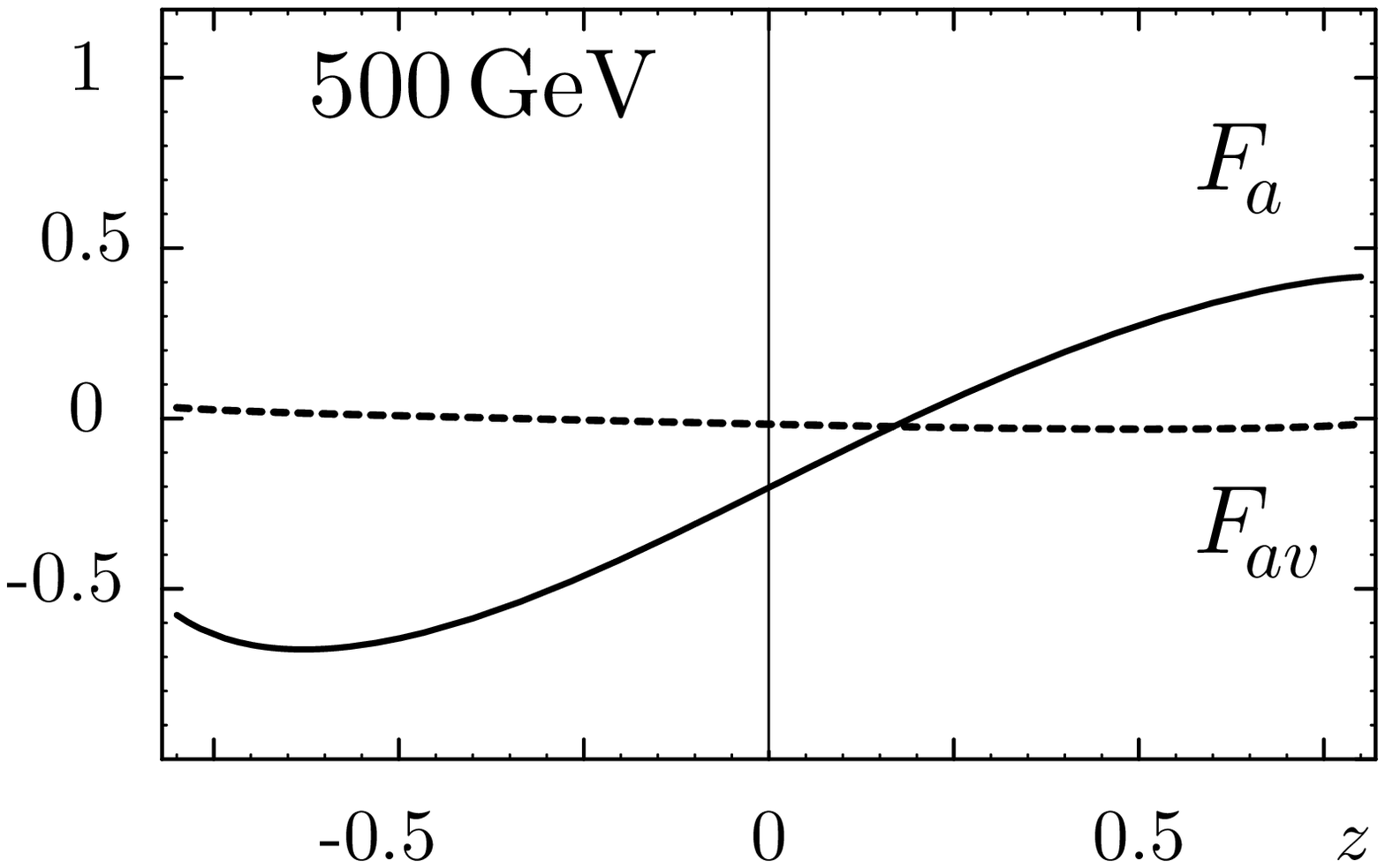}
\caption{Factors $F_{a}(\sqrt{s},z)$ (solid) and
$F_{av}(\sqrt{s},z)$ (dashed) in the normalized deviation of the
differential cross-section $d\tilde\sigma/dz$
for $\sqrt{s}=200$ and 500 GeV.}
\end{figure}
Now these factors are finite at $z\to 1$. Each of them in a
special way influences the differential cross-section.
\begin{enumerate}
\item
The factor at $\bar{v}^2$ is just the unity. Hence, the
four-fermion contact coupling between vector currents,
$\bar{v}^2$, determines the level of the deviation from the SM
value.
\item
The factor at $\bar{a}^2$ depends on the scattering angle in a
non-trivial way. It allows  to recognize the Abelian $Z'$ boson,
if the experimental accuracy is sufficient.
\item
The factor at $\bar{a}\bar{ v}$ results in small corrections.
\end{enumerate}

Thus, effectively, the obtained normalized differential
cross-section is a two-parametric function. In the next sections
we introduce the observables to fit separately each of these
parameters.

\subsection{Observables to pick out $\bar{v}^2$}

To recognize the signal of the Abelian $Z'$ boson by analyzing the
Bhabha process the differential cross-section deviation from the
SM predictions should be measured with a good accuracy. At
present, no such deviations have been detected at more than the
1$\sigma$ CL. In this situation it is resonable to introduce
integrated observables allowing to pick out $Z'$ signals by using
the most effective treating of available data. The observables
should be sensitive to the separate $Z'$ couplings. This admits of
searching for the $Z'$ signals in different processes as well as
to perform global fits.

The normalized deviation of the differential cross-section
(\ref{ncs}) is (effectively) the function of two parameters,
$\bar{a}^2$ and $\bar{v}^2$. We are going to introduce the
integrated observables which determine  separately the
four-fermion couplings $\bar{a}^2$ and $\bar{v}^2$ \cite{PRD2004}.

Let us first proceed with the observable for $\bar{v}^2$. After
normalization the factor at the vector-vector four-fermion
coupling becomes the unity. Whereas the factor at $\bar{a}^2$ is a
sign-varying function of the cosine of the scattering angle. As it
follows from Fig. 5, for the center-of-mass energy 200 GeV it is
small over the backward scattering angles. So, to measure the
value of $\bar{v}^2$ the normalized deviation of the differential
cross-section has to be integrated over the backward angles. For
the center-of-mass energy 500 GeV the factor at $\bar{a}^2$ is
already a non-vanishing quantity for the backward scattering
angles. The curves corresponding to intermediate energies are
distributed in between two these curves. Since they are
sign-varying ones at each energy point some interval of $z$ can be
chosen to make the integral to be zero. Thus, to measure the $Z'$
coupling to the electron vector current $\bar{v}^2$ we introduce
the integrated cross-section (\ref{ncs})
\begin{equation}\label{vobs}
\sigma_V = \int_{z_0}^{z_0+\Delta z} (d\tilde\sigma/dz)dz,
\end{equation}
where at each energy the most effective interval $[z_0,z_0+\Delta
z]$ is determined by the following requirements:
\begin{enumerate}
\item The relative contribution of the coupling $\bar{v}^2$ is
maximal. Equivalently, the contribution of the factor at
$\bar{a}^2$ is suppressed. \item The length $\Delta z$ of the
interval is maximal. This condition ensures that the largest
number of bins is taken into consideration.
\end{enumerate}

\begin{figure}
\centering
\includegraphics[bb= 91 3 322 234,width=65mm]{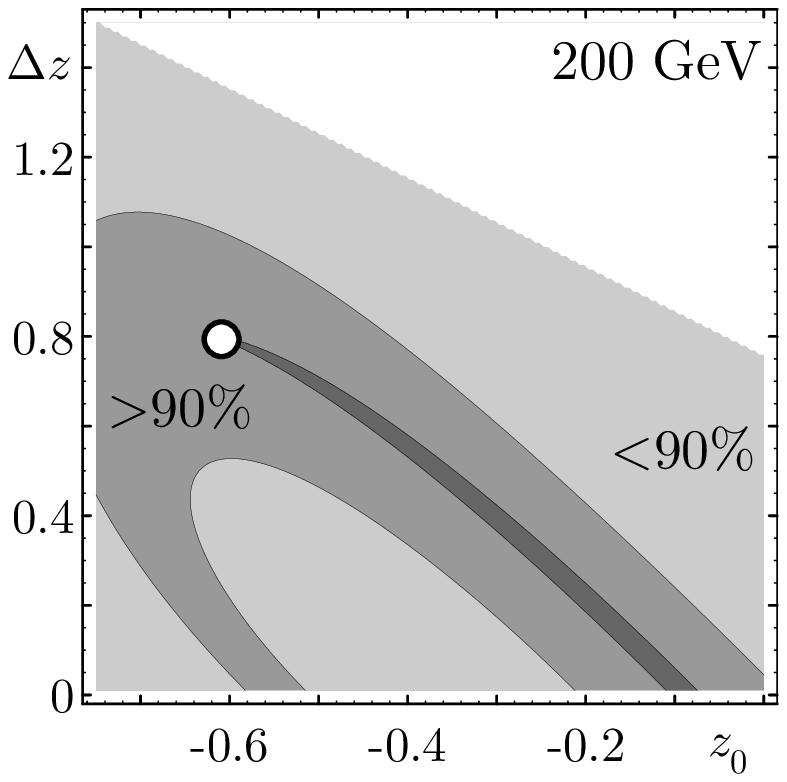}
\\
\includegraphics[bb= 91 3 322 234,width=65mm]{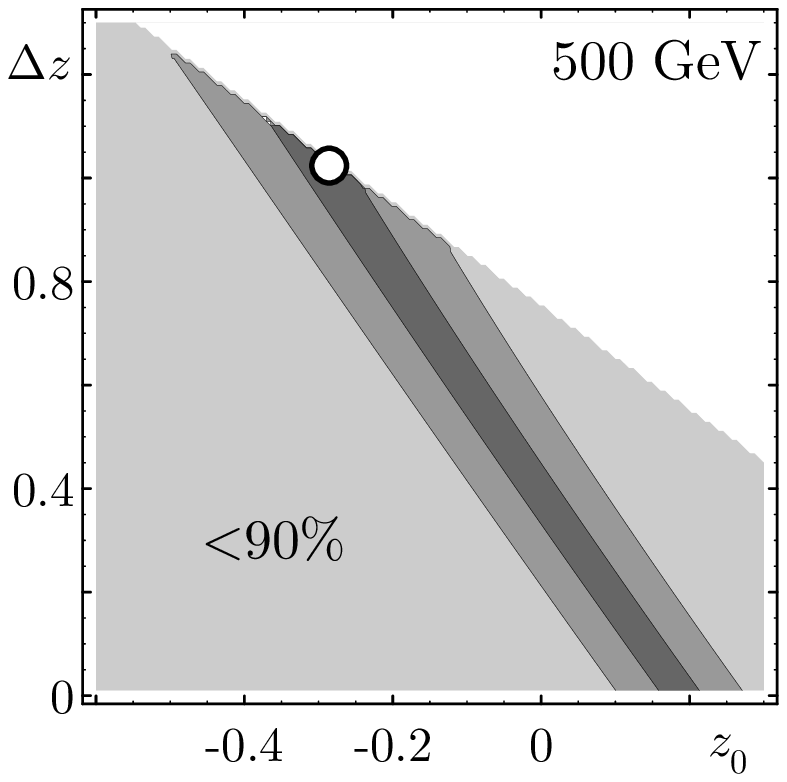}
\caption{Relative contribution of the factor at $\bar{v}^2$ to the
observable $\sigma_V$ as the function of the left boundary of the
angle interval, $z_0$, and the interval length, $\Delta z$, at the
center-of-mass energy 200 and 500 GeV. The shaded areas correspond
to the contributions $>95\%$
(dark), from 90\% to 95\% (midtone), and $<90\%$ (light).}
\end{figure}

The relative contribution of the factor at $\bar{v}^2$ is defined
as
\begin{equation}
\kappa_V= \frac{\Delta z}{\Delta z+\left|\int_{z_0}^{z_0+\Delta
z}F_a \,dz\right|+\left|\int_{z_0}^{z_0+\Delta
z}F_{av}\,dz\right|}
\end{equation}
and shown in Fig. 6 as the function of the left boundary of the
angle interval, $z_0$, and the interval length, $\Delta z$. In
each plot the dark area corresponds to the observables which
values are determined by the vector-vector coupling $\bar{v}^2$
with the accuracy $>95\%$. The area reflects the correlation of
the width of the integration interval $\Delta z$ with the choice
of the initial $z_0$ following from the mentioned requirements.
Within this area we choose the observable which includes the
largest number of bins (largest $\Delta z$). The corresponding
values of $z_0$ and $\Delta z$ are marked by the white dot on the
plots in Fig. 6. As the carried out analysis showed, the point
$z_0$ is shifted to the right with increase in energy whereas
$\Delta z$ remains approximately the same.

From the plots it follows that the most efficient intervals are
\begin{eqnarray}
&& -0.6<z<0.2,\quad\sqrt{s}=200\mbox{ GeV}, \nonumber\\&&
-0.3<z<0.7,\quad\sqrt{s}=500\mbox{ GeV}.
\end{eqnarray}
Therefore the observable (\ref{vobs}) allows to measure the $Z'$
coupling to the electron vector current $\bar{v}^2$ with the
efficiency $>95\%$.

Fitting the LEP2 final data with the one-parameter observable, we
find the values of the $Z'$ coupling to the electron vector
current together with their 1$\sigma$ uncertainties:
\begin{eqnarray}
  \mathrm{ALEPH}: &\bar{v}_e^2=& -0.11\pm 6.53 \times 10^{-4}\nonumber\\
  \mathrm{DELPHI}: &\bar{v}_e^2=& 1.60\pm 1.46 \times 10^{-4} \nonumber\\
  \mathrm{L3}: &\bar{v}_e^2=& 5.42\pm 3.72 \times 10^{-4} \nonumber\\
  \mathrm{OPAL}: &\bar{v}_e^2=& 2.42\pm 1.27 \times 10^{-4} \nonumber\\
  \mathrm{Combined}: &\bar{v}_e^2=& 2.24\pm 0.92 \times 10^{-4}. \nonumber
\end{eqnarray}
As one can see, the most precise data of DELPHI and OPAL
collaborations are resulted in the Abelian $Z'$ hints at one and
two standard deviation level, correspondingly. The combined value
shows the 2$\sigma$ hint, which corresponds to $0.006\le
|\bar{v}_e|\le 0.020$.

\subsection{Observables to pick out $\bar{a}^2$}

In order to pick the axial-vector coupling $\bar{a}^2$ one needs
to eliminate the dominant contribution coming from $\bar{v}^2$.
Since the factor at $\bar{v}^2$ in the $d\tilde\sigma/dz$  equals
unity, this can be done by summing up equal number of bins with
positive and negative weights. In particular, the forward-backward
normalized deviation of the differential cross-section appears to
be sensitive mainly to $\bar{a}^2$,
\begin{eqnarray}
\tilde\sigma_{\rm FB}&=&\int\nolimits_{0}^{z_{\rm
max}}dz\,\frac{d\tilde\sigma}{dz} -\int\nolimits_{-z_{\rm
max}}^{0}dz\,\frac{d\tilde\sigma}{dz} \nonumber\\ &=&
\tilde{F}_{a,\rm FB} \bar{a}^2 + \tilde{F}_{av,\rm
FB}\bar{a}\bar{v}.
\end{eqnarray}
The value $ z_{\rm max}$ is determined by the number of bins
included and, in fact, depends on the data set considered. The
LEP2 experiment accepted $e^+e^-$ events with $|z|<0.72$. In what
follows we take the angular cut $z_{\mathrm{max}}=0.7$ for
definiteness.

The efficiency of the observable is determined as:
\begin{equation}
\kappa= \frac{|\tilde{F}_{a,\rm FB}|}{|\tilde{F}_{a,\rm
FB}|+|\tilde{F}_{av,\rm FB}|}.
\end{equation}
It can be estimated as $\kappa=0.9028$ for the center-of-mass
energy 200 GeV and $\kappa=0.9587$ for 500 GeV. Thus, the
observable
\begin{eqnarray}\label{aobs}
&& \tilde\sigma_{\rm FB}=0.224 \bar{a}^2 - 0.024
\bar{a}\bar{v},\quad\sqrt{s}=200\mbox{ GeV}, \nonumber\\&&
\tilde\sigma_{\rm FB}=0.472 \bar{a}^2 - 0.020
\bar{a}\bar{v},\quad\sqrt{s}=500\mbox{ GeV}
\end{eqnarray}
is mainly sensitive to the $Z'$ coupling to the axial-vector
current $\bar{a}^2$.

Consider a usual situation when experiment is not able to
recognize the angular dependence of the differential cross-section
deviation from its SM value with the proper accuracy because of
loss of statistics. Nevertheless, a unique signal of the Abelian
$Z'$ boson can be determined. For this purpose the observables
$\int_{z_0}^{z_0+\Delta z}(d\tilde\sigma/dz)dz$ and
$\tilde\sigma_{\rm FB}$ must be measured. Actually, they are
derived from the normalized deviation of the differential
cross-section. If the deviation is inspired by the Abelian $Z'$
boson both the observables are to be positive quantities
simultaneously. This feature serves as the distinguishable signal
of the Abelian $Z'$ virtual state in the Bhabha process for the
LEP2 energies as well as for the energies of future
electron-positron collider ILC ($\geq 500$ GeV). The observables
fix the unknown low energy vector and axial-vector $Z'$ couplings
to the electron current. Their values have to be correlated with
the bounds on $\bar{a}^2$ and $\bar{v}^2$ derived by means of
independent fits for other scattering processes.

We estimated the observable (\ref{aobs}) related to the value of
$\bar{a}^2$. Since in the Bhabha process the effects of the
axial-vector coupling are suppressed with respect to those of the
vector coupling, we expect much larger experimental  uncertainties
for $\bar{a}^2$. Indeed, the LEP2 data lead to the huge errors for
$\bar{a}^2$ of order $10^{-3}-10^{-4}$
The mean values are negative numbers
which are too large to be interpreted as a manifestation of some
heavy virtual state beyond the energy scale of the SM.

Thus, the LEP2 data constrain the value of $\bar{v}^2$ at the
$2\sigma$ CL which could correspond to the Abelian $Z'$ boson with
the mass of the order 1 TeV. In contrast, the value of $\bar{a}^2$
is a large negative number with a significant experimental
uncertainty. This can not be interpreted as a manifestation of
some heavy virtual state beyond the energy scale of the SM.

\subsection{Many-parameter fits}

As the basic observable to fit the LEP2 experiment data on the
Bhabha process we propose the differential cross-section
\begin{equation}\label{7}
\left.\frac{d\sigma^\mathrm{Bhabha}}{dz}-\frac{d\sigma^{\mathrm{Bhabha},SM}}{dz}\right|_{z=z_i,\sqrt{s}=\sqrt{s_i}},
\end{equation}
where $i$ runs over the bins at various center-of-mass energies
$\sqrt{s}$. The final differential cross-sections measured by the
ALEPH (130-183 GeV, \cite{ALEPH}), DELPHI (189-207 GeV,
\cite{DELPHI}), L3 (183-189 GeV, \cite{L3}), and OPAL (130-207
GeV, \cite{OPAL}) collaborations are taken into consideration (299
bins).

As the observables for $e^+e^-\to\mu^+\mu^-,\tau^+\tau^-$
processes, we consider the total cross-section and the
forward-backward asymmetry
\begin{equation}\label{8}
\sigma^{\ell^+\ell^-}_T-\sigma_T^{\ell^+\ell^-,\mathrm{SM}},
\quad
\left.A^{\ell^+\ell^-}_{FB}-A_{FB}^{\ell^+\ell^-,\mathrm{SM}}\right|_{\sqrt{s}=\sqrt{s_i}},
\end{equation}
where $i$ runs over 12 center-of-mass energies $\sqrt{s}$ from 130
to 207 GeV. We consider the combined LEP2 data \cite{EWWG} for
these observables (24 data entries for each process). These data
are more precise as the corresponding differential cross-sections.
Our analysis is based on the fact that the kinematics of
$s$-channel processes is rather simple and the differential
cross-section is effectively a two-parametric function of the
scattering angle. The total cross-section and the forward-backward
asymmetry incorporate complete information about the kinematics of
the process and therefore are an adequate alternative for the
differential cross-sections.

The data are analysed by means of the $\chi^2$ fit \cite{PRD2007}.
Denoting the observables (\ref{7})--(\ref{8}) by $\sigma_i$, one
can construct the $\chi^2$-function,
\begin{equation}\label{9}
\chi^2(\bar{a}, \bar{v}_e,\bar{v}_\mu,\bar{v}_\tau) =
\sum\limits_i
\left[\frac{\sigma^\mathrm{ex}_i-\sigma^\mathrm{th}_i(\bar{a},
\bar{v}_e,\bar{v}_\mu,\bar{v}_\tau)}{\delta\sigma_i}\right]^{2},
\end{equation}
where $\sigma^\mathrm{ex}$ and $\delta\sigma$ are the experimental
values and the uncertainties of the observables, and
$\sigma^\mathrm{th}$ are their theoretical expressions presented
in Eqs. (\ref{4})--(\ref{5}). The sum in Eq. (\ref{9}) refers to
either the data for one specific process or the combined data for
several processes. By minimizing the $\chi^2$-function, the
maximal-likelihood estimate for the $Z'$ couplings can be derived.
The $\chi^2$-function is also used to plot the confidence area in
the space of parameters $\bar{a}$, $\bar{v}_e$, $\bar{v}_\mu$, and
$\bar{v}_\tau$. Note that in this way of experimental data
treating all the possible correlations are neglected. We believe
that at the present stage of investigation this is reasonable,
because the Collaborations have never reported on this
possibility.

For all the considered processes, the theoretic predictions
$\sigma^\mathrm{th}_i$ are linear combinations of products of two
$Z'$ couplings
\begin{eqnarray}\label{10}
\sigma^\mathrm{th}_i&=&\sum_{j=1}^{7} C_{ij}A_j,\\
 A_j&=&\{\bar{a}^2,\bar{v}_e^2,\bar{a}\bar{v}_e,\bar{v}_e\bar{v}_\mu,\bar{v}_e\bar{v}_\tau,
\bar{a}\bar{v}_\mu,\bar{a}\bar{v}_\tau\},
\nonumber
\end{eqnarray}
where $C_{ij}$ are known numbers. In what follows we use the
matrix notation $\sigma^\mathrm{th}=\sigma^\mathrm{th}_i$,
$\sigma^\mathrm{ex}=\sigma^\mathrm{ex}_i$, $C=C_{ij}$, $A=A_j$.
The uncertainties $\delta\sigma_i$ can be substituted by a
covariance matrix $D$. The diagonal elements of $D$ are
experimental errors squared,
$D_{ii}=(\delta\sigma^\mathrm{ex}_i)^2$, whereas the non-diagonal
elements are responsible for the possible correlations of
observables. The $\chi^2$-function can be rewritten as
\begin{eqnarray}\label{11}
\chi^2(A)&=&(\sigma^\mathrm{ex}-\sigma^\mathrm{th})^\mathrm{T}
D^{-1}
(\sigma^\mathrm{ex}-\sigma^\mathrm{th})
\nonumber\\
&=&(\sigma^\mathrm{ex}-CA)^\mathrm{T}
D^{-1} (\sigma^\mathrm{ex}-CA),
\end{eqnarray}
where the upperscript T denotes the matrix  transposition.

The $\chi^2$-function has a minimum, $\chi^2_\mathrm{min}$, at
\begin{equation}\label{12}
\hat{A}=(C^\mathrm{T}D^{-1}C)^{-1}C^\mathrm{T}D^{-1}\sigma^\mathrm{ex}
\end{equation}
corresponding to the maximum-likelihood values of $Z'$ couplings.
From Eqs. (\ref{11}), (\ref{12}) we obtain
\begin{eqnarray}\label{13}
\chi^2(A)-\chi^2_\mathrm{min}&=& (\hat{A}-A)^\mathrm{T}
\hat{D}^{-1}(\hat{A}-A),
\nonumber\\
\hat{D}&=& (C^\mathrm{T}D^{-1}C)^{-1}.
\end{eqnarray}

Usually, the experimental values $\sigma^\mathrm{ex}$ are
normal-distributed quantities with the mean values
$\sigma^\mathrm{th}$ and the covariance matrix $D$. The quantities
$\hat{A}$, being the superposition of $\sigma^\mathrm{ex}$, also
have the same distribution. It is easy to show that $\hat{A}$ has
the mean values $A$ and the covariance matrix $\hat{D}$.

The inverse matrix $\hat{D}^{-1}$ is symmetric and can be
diagonalized. The number of non-zero eigenvalues is determined by
the rank (denoted $M$) of $\hat{D}^{-1}$. The rank $M$ equals to
the number of linear-independent terms in the observables
$\sigma^\mathrm{th}$. So, the right-hand-side of Eq. (\ref{13}) is
a quantity distributed as $\chi^2$ with $M$ degrees of freedom
(d.o.f.). Since this random value is independent of $A$, the
confidence area in the parameter space ($\bar{a}$, $\bar{v}_e$,
$\bar{v}_\mu$, $\bar{v}_\tau$) corresponding to the probability
$\beta$ can be defined as \cite{SMEP}:
\begin{equation}\label{14}
\chi^2\le \chi^2_\mathrm{min}+\chi^2_{\mathrm{CL},\beta}(M),
\end{equation}
where $\chi^2_{\mathrm{CL,\beta}}(M)$ is the $\beta$-level of the
$\chi^2$-distribution with $M$ d.o.f.

In the Bhabha process, the $Z'$ effects are determined by 3
linear-independent contributions coming from $\bar{a}^2$,
$\bar{v}_e^2$, and $\bar{a}\bar{v}_e$ ($M=3$). As for the
$e^+e^-\to\mu^+\mu^-,\tau^+\tau^-$ processes, the observables
depend on 4 linear-independent terms for each process:
$\bar{a}^2$, $\bar{v}_e\bar{v}_\mu$, $\bar{v}_e\bar{a}$,
$\bar{a}\bar{v}_\mu$ for $e^+e^-\to\mu^+\mu^-$; and $\bar{a}^2$,
$\bar{v}_e\bar{v}_\tau$, $\bar{v}_e\bar{a}$, $\bar{a}\bar{v}_\tau$
for $e^+e^-\to\tau^+\tau^-$ ($M=4$). Note that some terms in the
observables for different processes are the same. Therefore, the
number of $\chi^2$ d.o.f. in the combined fits is less than the
sum of d.o.f. for separate processes. Hence, the predictive power
of the larger set of data is not drastically spoiled by the
increased number of d.o.f. In fact, combining the data of the
Bhabha and $e^+e^-\to\mu^+\mu^-$ ($\tau^+\tau^-$) processes
together we have to treat 5 linear-independent terms. The complete
data set for all the lepton processes is ruled by 7 d.o.f. As a
consequence, the combination of the data for all the lepton
processes is possible.

The parametric space of couplings ($\bar{a}$, $\bar{v}_e$,
$\bar{v}_\mu$, $\bar{v}_\tau$) is four-dimensional. However, for
the Bhabha process it is reduced to the plane ($\bar{a}$,
$\bar{v}_e$), and to the three-dimensional volumes ($\bar{a}$,
$\bar{v}_e$, $\bar{v}_\mu$), ($\bar{a}$, $\bar{v}_e$,
$\bar{v}_\tau$) for the $e^+e^-\to\mu^+\mu^-$ and
$e^+e^-\to\tau^+\tau^-$ processes, correspondingly. The predictive
power of data is distributed not uniformly over the parameters.
The parameters $\bar{a}$ and $\bar{v}_e$ are present in all the
considered processes and appear to be significantly constrained.
The couplings $\bar{v}_\mu$ or $\bar{v}_\tau$ enter when the
processes $e^+e^-\to\mu^+\mu^-$ or $e^+e^-\to\tau^+\tau^-$ are
accounted for. So, in these processes, we also study the
projection of the confidence area onto the plane
($\bar{a},\bar{v}_e$).

The origin of the parametric space, $\bar{a}=\bar{v}_e=0$,
corresponds to the absence of the $Z'$ signal. This is the SM
value of the observables. This point could occur inside or outside
of the confidence area at a fixed CL. When it lays out of the
confidence area, this means the distinct signal of the Abelian
$Z'$. Then the signal probability can be defined as the
probability that the data agree with the Abelian $Z'$ boson
existence and exclude the SM value. This probability corresponds
to the most stringent CL (the largest $\chi^2_\mathrm{CL}$) at
which the point $\bar{a}=\bar{v}_e=0$ is excluded. If the SM value
is inside the confidence area, the $Z'$ boson is indistinguishable
from the SM. In this case, upper bounds on the $Z'$ couplings can
be determined.

\begin{figure}
\centering
  \includegraphics[bb= 0 0 540 300, width=.4\textwidth]{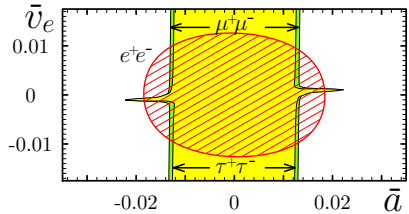}
  \caption{The 95\% CL areas in the ($\bar{a},\bar{v}_e$) plane for the Bhabha,
  $e^+e^-\to\mu^+\mu^-$, and $e^+e^-\to\tau^+\tau^-$ processes.}\label{fig:1}
\end{figure}

The 95\% CL areas in the ($\bar{a},\bar{v}_e$) plane for the
separate processes are plotted in Fig. \ref{fig:1}. As it is seen,
the Bhabha process constrains both the axial-vector and vector
couplings. As for the $e^+e^-\to\mu^+\mu^-$ and
$e^+e^-\to\tau^+\tau^-$ processes, the axial-vector coupling is
significantly constrained, only. The confidence areas include the
SM point at the meaningful CLs, so the experiment could not pick
out clearly the Abelian $Z'$ signal from the SM. An important
conclusion from these plots is that the experiment significantly
constrains only the couplings entering sign-definite terms in the
cross-sections.

\begin{figure}
\centering
  \includegraphics[bb= 0 0 540 470 ,width=.4\textwidth]{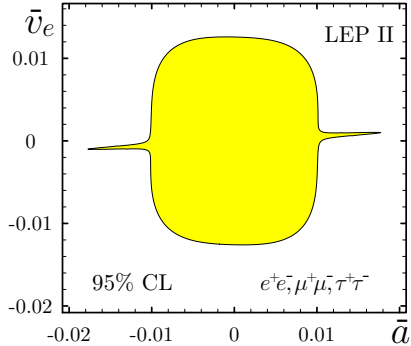}\\
  \caption{The projection of the 95\% CL area onto the ($\bar{a},\bar{v}_e$) plane
  for the combination of the Bhabha, $e^+e^-\to\mu^+\mu^-$, and $e^+e^-\to\tau^+\tau^-$
  processes.}\label{fig:2}
\end{figure}

The combination of all the lepton processes is presented in Fig.
\ref{fig:2}. There is no visible signal beyond the SM. The
couplings to the vector and axial-vector electron currents are
constrained by the many-parameter fit as $|\bar{v}_e|<0.013$,
$|\bar{a}|<0.019$ at the 95\% CL. If the charge corresponding to
the $Z'$ interactions is assumed to be of order of the
electromagnetic one, then the $Z'$ mass should be greater than
0.67 TeV. For the charge of order of the SM $SU(2)_L$ coupling
constant $m_{Z'}\ge 1.4$ TeV. One can see that the constraint is
not too severe to exclude the $Z'$ searches at the LHC.

\begin{figure}
\centering
  \includegraphics[bb= 0 0 540 390,width=.4\textwidth]{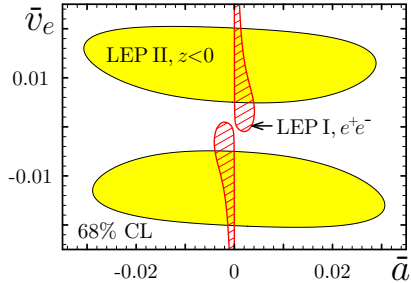}\\
  \caption{The 68\% CL area in the ($\bar{a},\bar{v}_e$) plane
  from the backward bins of the Bhabha process in the LEP2 experiments
  (the shaded area). The hatched area is the 68\% CL area from
  the LEP 1 data on the Bhabha process.}\label{fig:3}
\end{figure}

Let us compare the obtained results with the one-parameter fits.
As one can see, the most precise data of DELPHI and OPAL
collaborations are resulted in the Abelian $Z'$ hints at one and
two standard deviation level, correspondingly. The combined value
shows the 2$\sigma$ hint, which corresponds to $0.006\le
|\bar{v}_e|\le 0.020$. On the other hand, our many-parameter fit
constrains the $Z'$ coupling to the electron vector current as
$|\bar{v}_e|\le 0.013$ with no evident signal. Why does the
one-parameter fit of the Bhabha process show the 2$\sigma$ CL hint
whereas there is no signal in the two-parameter one? Our
one-parameter observable accounts mainly for the backward bins.
This is in accordance with the kinematic features of the process:
the backward bins depend mainly on the vector coupling
$\bar{v}^2_e$, whereas the contributions of other couplings are
kinematically suppressed (see Fig. 4). Therefore, the difference
of the results can be inspired by the data sets used. To clarify
this point, we perform the many-parameter fit with the 113
backward bins ($z\le 0$), only. The $\chi^2$ minimum,
$\chi^2_\mathrm{min}=93.0$, is found in the non-zero point
$|\bar{a}|=0.0005$, $\bar{v}_e= 0.015$. This value of the $Z'$
coupling $\bar{v}_e$ is in an excellent agreement with the mean
value obtained in the one-parameter fit. The 68\% confidence area
in the ($\bar{a},\bar{v}_e$) plane is plotted in Fig. \ref{fig:3}.
There is a visible hint of the Abelian $Z'$ boson. The zero point
$\bar{a}=\bar{v}_e=0$ (the absence of the $Z'$ boson) corresponds
to $\chi^2=97.7$. It is covered by the confidence area with
$1.3\sigma$ CL. Thus, the backward bins show the $1.3\sigma$ hint
of the Abelian $Z'$ boson in the many-parameter fit. So, the
many-parameter fit is less precise than the analysis of the
one-parameter observables.

At LEP1 experiments \cite{LEP1} the $Z$-boson couplings to the
vector and axial-vector lepton currents ($g_V$, $g_A$) were
precisely measured. The Bhabha process shows the 1$\sigma$
deviation from the SM values for Higgs boson masses $m_H\ge 114$
GeV (see Fig. 7.3 of Ref. \cite{LEP1}). This deviation could be
considered as the effect of the $Z$--$Z'$ mixing. It is
interesting to estimate the bounds on the $Z'$ couplings following
from these experiments.

Due to the RG relations, the $Z$--$Z'$ mixing angle is completely
determined by the axial-vector coupling $\bar{a}$. So, the
deviations of $g_V$, $g_A$ from their SM values are governed by
the couplings $\bar{a}$ and $\bar{v}_e$,
\begin{equation}\label{lep1}
g_V-g_V^{\mathrm{SM}}=-49.06 \bar{a}\bar{v}_e,\quad
g_A-g_A^{\mathrm{SM}} = 49.06 \bar{a}^2.
\end{equation}
Let us assume that the total deviation of theory from experiments
follows due to the $Z$--$Z'$ mixing. This gives an upper bound on
the $Z'$ couplings. In this way one can estimate whether  the $Z'$
boson is excluded by the experiments or not.

The 1$\sigma$ CL area for the Bhabha process from Ref. \cite{LEP1}
is converted into the ($\bar{a},\bar{v}_e$) plane in Fig.
\ref{fig:3}. The SM values of the couplings correspond to the top
quark mass $m_t=178$ GeV and the Higgs scalar mass $m_H=114$ GeV.
As it is seen, the LEP1 data on the Bhabha process is compatible
with the Abelian $Z'$ existence at the $1\sigma$ CL. The
axial-vector coupling is constrained as $|\bar{a}|\le 0.005$. This
bound corresponds to $\bar{a}^2\le 2.5\times 10^{-5}$, which
agrees with the one-parameter fits of the LEP2 data for
$e^+e^-\to\mu^+\mu^-,\tau^+\tau^-$ processes ($\bar{a}^2= 1.3\pm
3.89\times 10^{-5}$ at 68\% CL). On the other hand, the vector
coupling constant $\bar{v}_e$ is practically unconstrained by the
LEP1 experiments.

For the convenience, in Table 4 we collect the summary of the fits
of the LEP data in terms of dimensionless contact couplings
(\ref{6}). From the analysis carried out we come to conclusion
that, in principle, the LEP experiments were able to detect the
$Z'$-boson signals if the statistics had been sufficient.

\begin{table}
  \centering
  \caption{The summary of the fits of the LEP data for
  the dimensionless contact couplings (\ref{6}).}
\begin{tabular}{|l|l|l|}
  \hline
  Data & $\bar{v}^2_e$ & $\bar{a}^2$ \\
  \hline\hline
  \multicolumn{3}{|c|}{LEP1} \\
  \hline
  $e^-e^+$, 68\% CL & - & $(1.25\pm1.25)\times 10^{-5}$ 
  \\
  \hline
  \multicolumn{3}{|c|}{LEP2, one-parameter fits} \\
  \hline
  $e^-e^+$, 68\% CL & $(2.24\pm 0.92)\times 10^{-4}$ 
  & - \\
  $\mu\mu$, 68\% CL & - & $(3.66^{+4.89}_{-4.86})\times 10^{-5}$ 
  \\
  $\mu\mu$,$\tau\tau$, 68\% CL & - & $(1.33^{+3.89}_{-3.87})\times 10^{-5}$\\
  \hline
  \multicolumn{3}{|c|}{LEP2, many-parameter fits} \\
  \hline
  $e^-e^+,\mu\mu,\tau\tau$, 95\% CL & $\le 1.69\times 10^{-4}$ & $\le 3.61\times 10^{-4}$\\
  $e^-e^+$ backward, 68\% CL & $(2.25^{+1.79}_{-2.07})\times 10^{-4}$ & $\le 9.49\times 10^{-4}$\\
  \hline
\end{tabular}
\end{table}

\section{$Z'$ hints within neural network analysis}
Since the actual LEP2 data set is not too large to detect $Z'$
boson, one needs in the estimate of its parameters which could be
used in future experiments. To determine them in a maximally full
way we address to the analysis based on the predictions of the
neural networks (for applications in high energy physics see, for
example, \cite{dudko}). The main idea of this approach is to
constrain a given data set in such a way that an amount of it is
considered as an inessential background and omitted. The remaining
data are expected to give a more precise fit of the parameters of
interest.

We take into consideration the complete set of the differential
cross sections for the Bhabha process accumulated by all the LEP
Collaborations and apply the following criteria to restrict the
data \cite{Buryk}:
\begin{enumerate}
    \item
As the signal we use the differential cross sections for the
Bhabha process accounting for the SM plus $Z'$ and calculated at
$0.25 \times 10^{-4} \leq \bar{v}_e^2 \leq 4 \times 10^{-4}$ and
$0.25 \times 10^{-4} \leq \bar{a}_e^2 \leq 4 \times 10^{-4}$.

Such a choice of parameters is motivated by the results obtained
in the previous sections. The cross sections due to the $Z'$
exchange diagrams were calculated with the RG relations been taken
into consideration.
    \item
As the background we use the deviations of the experimental
differential cross sections from calculated for the SM plus $Z'$
ones which are larger than redoubled uncertainties of LEP2
experimental data.
\end{enumerate}

The network trained with these criteria omits the events which
correspond to the large deviations from the theoretical cross
sections but accounts for the peculiarities proper the $Z'$
existence. To construct and train the neural network we used the
program MLPFit \cite{MLPFit}. The results of the carried out
analysis based on the two parametric fit discussed in the previous
sections demonstrate the $2\sigma$ CL hint for the $Z'$.  For the
vector coupling the neural network at the $2\sigma $ CL predicts
$\bar{v}_e^2 = 2.4 \pm 1.99 \times 10^{- 4}$ \cite{Buryk} that is
in agreement with the discussed above result derived in the one
parametric fit. The obtained values of $\bar{v}_e^2$ correspond to
the value of the mass $m_{Z'} = 0.53 - 1.05 $ TeV, if the coupling
$\tilde{g}$ is of the order of the SM gauge couplings, $
g^2/(4\pi)\sim 0.01 - 0.03$. Thus, the carried out analysis
demonstrates the hint of $Z'$ boson which can be not too heavy. We
conclude once again that the data set of the LEP experiments is
not sufficient to detect the pronounced signal of this virtual
particle.

\section{Search for Chiral $Z'$ in Bhabha process}

Let us turn to the analysis of the Bhabha process with the aim to
search for the Chiral $Z'$ gauge boson \cite{YAF2007}. The Chiral
$Z'$ interacts with the SM doublets only that can be described by
one parameter for each doublet ($\tilde{Y}_{fL}$ and
$\tilde{Y}_\phi$). It is characterized by the constraints
(\ref{rgr1})
\begin{equation}\label{3}
\tilde{Y}_{L,f}=-\tilde{Y}_{fL}\,\sigma_3,\quad
\tilde{Y}_{R,f}=0,\quad
\tilde{Y}_{\phi_i}=-\tilde{Y}_{\phi}\,\sigma_3
\end{equation}
where $\sigma_3$ is the Pauli matrix.

Remind that in the Bhabha process it is convenient to use the
normalized cross-sections (\ref{ncs}):
\[
\frac{d\tilde\sigma}{dz}=\frac{m_Z^2}{{4\pi}f^{ee}_2(z)}\Delta
\frac{d\sigma}{dz} = \bar{v}_e^2 + F_a \bar{a}_e^2 + F_{av}
\bar{a}_e\bar{v}_e +F_{v\phi} \bar{v}_e\bar\phi +F_{a\phi}
\bar{a}_e\bar\phi.
\]
Since the Chiral $Z'$ boson does not interact with the
right-handed species, the normalized deviation of the differential
cross-section from its SM prediction is determined by two factors,
$F_L$ and $F_{L\phi}$,
\[
\frac{d\tilde\sigma}{dz}=\frac{m_Z^2}{{4\pi}f^{ee}_2(z)}\Delta
\frac{d\sigma}{dz} = F_L \bar{l}_e^2 + F_{L\phi}
\bar{l}_e\bar\phi,
\]
where we define the dimensionless constants
\[
\bar{l}_f=\frac{m_Z}{\sqrt{4\pi}m_{Z'}}\tilde{g}\tilde{Y}_{fL},\quad
\bar{\phi}=\frac{m_Z}{\sqrt{4\pi}m_{Z'}}\tilde{g}\tilde{Y}_\phi.
\]
The normalization gives us two benefits. First, the obtained
factors $F(\sqrt{s},z)$ are finite for all values of the
scattering angle $z$. Second, the experimental uncertainties for
different bins become equalized that provides the statistical
equivalence of different bins. The latter is important for the
construction of integrated cross-sections.

The $Z$-$Z'$ mixing angle is determined by $\bar\phi$ as follows,
\[
\theta_0\simeq\frac{m_W
\sin\theta_W}{\sqrt{\alpha_\mathrm{em}}m_{Z'}}\bar\phi,
\]
where $\alpha_\mathrm{em}$ is the fine structure constant.

\subsection{One-parameter fit for Chiral $Z'$}

The Chiral $Z'$ boson does not interact with the right-handed
species. The normalized deviation of the differential
cross-section from its SM prediction,
\[
d\tilde\sigma/dz = F_L \bar{l}_e^2 + F_{L\phi} \bar{l}_e\bar\phi,
\]
is determined by two finite factors, $F_L$ and $F_{L\phi}$, which
are shown in Fig. 10.
\begin{figure}\label{fig-dcs-ch-norm}
\centering
\includegraphics[bb= 0 0 300 200,width=.5\textwidth]{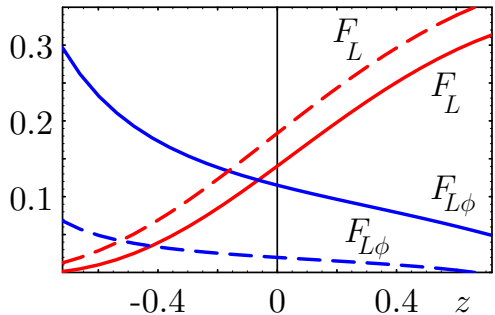}
\caption{ The normalized factors $F_L$ and $F_{L\phi}$ describing
the Chiral $Z'$ effects in the Bhabha process at $\sqrt{s}=200$
GeV (solid lines)  and $\sqrt{s}=500$ GeV (dashed lines).}
\end{figure}

As it is seen, the four-fermion contact coupling $\bar{l}_e^2$
contributes mainly to the forward scattering angles, whereas the
$Z$-$Z'$ mixing term affects the backward angles. At the LEP
energies they can be of the same order of magnitude. The
contribution of the mixing vanishes with the energy growth. We
have a possibility to derive the effective experimental
constraints   on them without any additional restrictions.

First, let us construct a one-parametric observable which is most
preferred by the statistical treatment of data. As is clear, it is
impossible to separate the couplings $\bar{l}_e^2$ and
$\bar{l}_e\bar\phi$ in any observable which is an integrated
cross-section over some interval of $z$. However, the mixing
contribution can be eliminated in the cross-section of the form
(which is inspired by the forward-backward asymmetry)
\[
\Delta\sigma(z^*) = \int_{z^*}^{z_\mathrm{max}}
\frac{d\tilde\sigma}{dz} \, dz - \int_{-z_\mathrm{max}}^{z^*}
\frac{d\tilde\sigma}{dz} \, dz,
\]
where the boundary value $z^*$ should be chosen to suppress the
coefficient at $\bar{l}_e\bar\phi$. The maximal value of the
scattering angle $z_\mathrm{max}$ is determined by a particular
experiment. In this way we introduce the one-parametric
sign-definite observable sensitive to $\bar{l}_e^2$.

The LEP Collaborations DELPHI and L3 measured the differential
cross-sections with $z_\mathrm{max}=0.72$ \cite{DELPHI,L3}. The
set of boundary angles $z^*$ as well as the theoretic and
experimental values of the observable are collected in Table 5.
The other LEP Collaborations -- ALEPH and OPAL -- used
$z_\mathrm{max}=0.9$ \cite{OPAL,ALEPH}. The corresponding data are
presented in Table 6.
\begin{table}\label{tab-1}
\centering \caption{The boundary angles $z^*$ and the theoretic
and experimental values of the observable $\Delta\sigma(z^*)$ at
$z_\mathrm{max}=0.72$.} \rule{0pt}{10pt}\\
\begin{tabular}{|c|c|c|r|r|} \hline $\sqrt{s}$,
GeV & $z^*$ & $\Delta\sigma(z^*)$ &
$\Delta\sigma^\mathrm{ex}(z^*)$, DELPHI &
$\Delta\sigma^\mathrm{ex}(z^*)$, L3 \\
\hline\hline
 183 & -0.245 & 1742 $\bar{l}_e^2$ &-& $-2.38 \pm 7.03$ \\
 189 & -0.252 & 1775 $\bar{l}_e^2$ & $-4.28 \pm 3.36$ & $3.05 \pm 3.65$\\
 192 & -0.255 & 1788 $\bar{l}_e^2$ & $4.57 \pm 7.32$ &-\\
 196 & -0.259 & 1806 $\bar{l}_e^2$ & $3.77 \pm 4.33$ &-\\
 200 & -0.263 & 1823 $\bar{l}_e^2$ & $-1.95 \pm 4.05$ &-\\
 202 & -0.265 & 1831 $\bar{l}_e^2$ & $1.31 \pm 5.54$ &-\\
 205 & -0.267 & 1843 $\bar{l}_e^2$ & $-4.09 \pm 3.89$ &-\\
 207 & -0.269 & 1851 $\bar{l}_e^2$ & $0.40 \pm 3.33$ &-\\
\hline
\end{tabular}
\end{table}
\begin{table}\label{tab-2}
\centering \caption{The boundary angles $z^*$ and the theoretic
and experimental values of the observable $\Delta\sigma(z^*)$ at
$z_\mathrm{max}=0.9$.}\rule{0pt}{10pt}\\
\begin{tabular}{|c|c|c|r|r|}
\hline $\sqrt{s}$, GeV & $z^*$ & $\Delta\sigma(z^*)$ &
$\Delta\sigma^\mathrm{ex}(z^*)$, ALEPH &
$\Delta\sigma^\mathrm{ex}(z^*)$, OPAL \\
\hline\hline
 130 & -0.217 & 2017 $\bar{l}_e^2$ & $-12.40 \pm 19.24$ & $-4.13 \pm 29.29$ \\
 136 & -0.266 & 2092 $\bar{l}_e^2$ & $-50.21 \pm 16.64$ & $-34.18 \pm 31.58$ \\
 161 & -0.370 & 2311 $\bar{l}_e^2$ & $-15.90 \pm 13.24$ & $-14.02 \pm 22.32$ \\
 172 & -0.400 & 2398 $\bar{l}_e^2$ & $-12.11 \pm 12.50$ & $13.71 \pm 17.84$ \\
 183 & -0.424 & 2474 $\bar{l}_e^2$ & $-1.51 \pm ~5.18$ & $11.04 \pm ~5.57$ \\
 189 & -0.435 & 2512 $\bar{l}_e^2$ &-& $-0.63 \pm ~3.28$ \\
 192 & -0.441 & 2531 $\bar{l}_e^2$ &-& $-3.48 \pm 9.85$ \\
 196 & -0.447 & 2554 $\bar{l}_e^2$ &-& $2.96 \pm ~5.09$ \\
 200 & -0.454 & 2577 $\bar{l}_e^2$ &-& $0.35 \pm ~4.68$ \\
 202 & -0.457 & 2587 $\bar{l}_e^2$ &-& $-2.87 \pm ~9.00$ \\
 205 & -0.461 & 2604 $\bar{l}_e^2$ &-& $5.88 \pm ~4.67$ \\
 207 & -0.464 & 2614 $\bar{l}_e^2$ &-& $-1.42 \pm ~3.46$ \\
\hline
\end{tabular}
\end{table}

The standard $\chi^2$-fit gives the following constraints for the
coupling $\bar{l}_e^2$ at the 68\% CL:
\begin{eqnarray}
  \mathrm{ALEPH:} && \bar{l}_e^2= -0.00304\pm 0.00176 \nonumber\\
  \mathrm{DELPHI:} && \bar{l}_e^2= -0.00054\pm 0.00086\nonumber\\
  \mathrm{L3:} && \bar{l}_e^2= 0.00109\pm 0.00184 \nonumber\\
  \mathrm{OPAL:} && \bar{l}_e^2= 0.00051\pm 0.00064\nonumber\\
  \mathrm{Combined:} && \bar{l}_e^2= -0.00004\pm 0.00048\nonumber
\end{eqnarray}
Hence it is seen that the most precise data of DELPHI and OPAL
collaborations give no signal of the Chiral $Z'$ at the 1$\sigma$
CL. The combined value also shows no signal at the 1$\sigma$ CL.
From the combined fit the 95\% CL bound on the value of
$\bar{l}_e^2$ can be derived, $\bar{l}_e^2<9\times 10^{-4}$.
Supposing the $Z'$ coupling constant $\tilde{g}$ to be of the
order of the electroweak one, $\tilde{g}\simeq 0.6$, the
corresponding $Z'$ mass has to be larger than $0.5$ TeV.

\subsection{Two parametric fit for  Chiral $Z'$}

A complete two parametric fit of experimental data based directly
on the differential cross-sections has been carried out in Ref.
\cite{YAF2007}. Due to only two independent couplings this fit is
efficient. In the fitting  the available final data for the
differential cross-sections of the Bhabha process were used. The
data set consists of 299 bins including the data of ALEPH at
130-183 GeV, DELPHI at 189-207 GeV, L3 at 183-189 GeV, and OPAL at
130-207 GeV \cite{DELPHI,OPAL,ALEPH,L3}. The fitting procedure is
similar to that of discussed above for the Abelian $Z'$. The
results can be summarized as follows.

The parameter space of the Chiral $Z'$ is the plane ($\bar{l}_e$,
$\bar\phi$). The minimum of the $\chi^2$-function,
$\chi^2_\mathrm{min}=237.29$, is reached at zero value of
$\bar{l}_e$ ($\simeq 10^{-4}$) and almost independent of the value
of $\bar\phi$ (the maximal-likelihood values of the couplings).
The 95\% CL area ($\chi^2_\mathrm{CL}=5.99$) is shown in Figure 3.
\begin{figure}\label{fig-ch-many}
\centering
\includegraphics[bb= 0 0 275 235,width=.5\textwidth]
{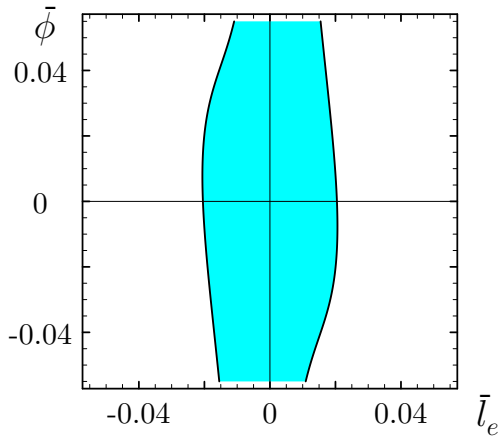} \caption{The 95\% CL area in $\bar{l}_e-\bar\phi$
plane. The final data of ALEPH 130-183 GeV, DELPHI GeV 189-207, L3
183-189 GeV, and OPAL 130-207 GeV are combined.}
\end{figure}

As one can see, the zero point, $\bar{l}_e = \bar\phi= 0$ (the
absence of the Chiral $Z'$ boson) is inside the confidence area.
The value of $\chi^2$ in this point (238.62) is indistinguishable
from the $\chi^2_\mathrm{min}$. In other words, the set of
experimental data cannot determine the signal of the Chiral
$Z'$-boson.

As also is seen, the value of $\bar{l}_e$ is constrained as
$\bar{l}_e<0.02$ at the 95\% CL. This upper bound is in an
agreement with the corresponding result of the one-parameter fit
($\bar{l}_e<0.03$). Thus, the $Z'$ mass has to be larger than
$0.75$ TeV, if the $Z'$ coupling constant $\tilde{g}$ is again
supposed to be of the order of the electroweak one,
$\tilde{g}\simeq 0.6$.

The fit of the differential cross-sections leads to a better
accuracy for $\bar{l}_e$ than the fit of the integrated
cross-sections based on  the same data. Without accounting for the
model-independent relations between the $Z'$ couplings it is
impossible to obtain such results.

\section{Model independent results and search for $Z'$ at the LHC}

In this section we  discuss  all the assumptions giving a
possibility to pick out the $Z'$ signal and determine its
characteristics in a model independent way. We also note the role
of the present results for the future LHC and ILC experiments. As
it was already stressed, in searching for this particle at the LEP
and Tevatron a model dependent analysis was mainly used. The
motivation for this was  the different number of chiral fermions
involved in different models (see, for example, \cite{Lang08}). In
this approach, the low bounds on $m_{Z'}$ have been estimated and
a smallness of the $Z - Z'$ mixing was also observed.

On the contrary, in our model independent approach the RG
relations between the parameters of the effective low energy
Lagrangians have been accounted for that gave a possibility to
determine not only the bounds but also the mass and other
parameters of the $Z'$.

First we  note all the assumptions used in our considerations. We
analyzed the four-fermion scattering amplitudes of order $\sim
m_{Z'}^{-2}$ generated by the $Z'$ virtual states. The vertices
linear in $Z'$ were included into the effective low-energy
Lagrangian. We also impose a number of natural conditions. The
interactions of a renormalizable type are dominant at low energies
$\sim m_W$. The non-renormalizable interactions generated at high
energies due to radiation corrections are suppressed by the
inverse heavy mass $\sim 1/m_{Z'}$ and neglected. We also assumed
that the $SU(2)_L \times U(1)_Y$ gauge group of the SM is a
subgroup of the GUT group. As a consequence, all structure
constants connecting two SM gauge bosons with $Z'$ are to be zero.
Hence, the interactions of gauge fields of the types $Z' W^+ W^-,
Z' Z Z$, and other are absent at the tree level. Our effective
Lagrangian is also consistent with the absence  of the tree-level
flavor-changing neutral currents (FCNCs) in the fermion sector.
The renormalizable interactions of fermions and scalars are
described by the Yukawa Lagrangian (\ref{Lyukawa}) that accounts
for different possibilities of the Yukawa sector without the
tree-level FCNCs.  These assumptions are quite general and
satisfied in a wide class of $E_6$ inspired models.

Within these constraints for the low energy effective Lagrangian
the RG  correlations have been derived. Correspondingly, the model
independent estimates of the mass $m_{Z'}$ and other parameters
are regulated by  the noted requirements.  Therefore, the extended
underlying model has also to accept  them.

In this regard, let us discuss the role of the obtained estimates
for the LHC. As it is well known (see, for example,
\cite{Lang08,Rizzo06}), there are many tools at the LHC for $Z'$
identification. But many of them are only applicable if this
particle is relatively light. Our results are in favor to this
case.

Next important point is the determination of $Z'$ couplings to the
various SM fermions. As we have shown, the axial-vector couplings
of the $Z'$ to the SM fermions are universal and proportional to
its coupling to the Higgs field. Hence we have obtained an
estimate of the $a = a_f$ couplings for both leptons and quarks.
This is an essential input because experimental analyses for the
LHC have mainly concentrated on being able to distinguish models
and not on actual couplings. The vector coupling $v_e$ was also
estimated that, in particular,  may help to distinguish the decay
of the $Z'$ resonance state to $  e^+ e^-$ pairs. Since the
couplings $a_e$ and $v_e$ were estimated there is a possibility to
distinguish this process from the decay of the $K K $ system. In
the literature \cite{Rizzo06,Dittmar}  on searching for the $Z'$
it is also mentioned  that the determination of the $Z'$ couplings
to fermions could be fulfilled channel by channel, $a_e, v_e,
v_{e,b}, a_{e,b}$, \ldots. In  almost all these considerations the
relations between the parameters  have not been taken into
account. But this is very essential for  treating of experimental
data and introducing  relevant observables to measure.

Other  parameter is the $Z - Z'$ mixing, which is responsible for
different decay processes and the effective interaction vertices
generated at the LHC \cite{Lang08,Rizzo06}. It is also determined
by the axial-vector coupling (see Eq. (\ref{grgav})) and estimated
in a model independent way. Remind that in our analysis (in
contrast to the approaches of the LEP Collaborations) the mixing
was systematically taken into account. Its value is of the same
order of magnitude as the parameters that were fitted in
experiments. Note also that the existence of other heavy particles
with  masses $m_X \ge m_{Z'}$ does not influence the RG relations
which are the consequences of the necessary condition for
renormalizability. In fact, this condition (the structure of a
divergence generated by radiation corrections coincides with that
of the tree-level vertex)  holds for each renormalizable type
interaction.

An important role of the model independent results for searching
for  $Z'$ at the  LHC and ILC consists, in particular, in
possibility to determine the particle as a virtual state due to a
large amount of relevant events. We mentioned already that, in
principle, LEP experiments were able to determine it if the
statistics was sufficiently large. Experiments at the ILC will
increase numerously the data set of interest. In fact, the
observables, introduced in sects. 6 and 7 for picking out uniquely
$a_f^2$ and $v_e^2$ couplings in the leptonic scattering process,
are also effective at energies $\sqrt{s}\ge 500$ GeV and could be
applied in future experiments at ILC.

Other model independent methods of searching for the $Z'$ as a
resonance state are proposed in the literature (see Refs.
\cite{Dittmar,zplhc1,zplhc2}). We do not discuss them here because
they take no relations between the parameters into consideration.
Besides, the main goal of the present paper is to adduce model
independent information about the $Z'$ followed from experiments
at low energies. Different aspects of $Z'$ physics at the LHC are
out of the scope of it.

\section{Discussion}

In this section  we collect in a convenient form all the results
obtained  and make a comparison with other investigations on
searching  for $Z'$ at low energies. In fact, this is a large area
to discuss. References to numerous results obtained in either
model dependent or model independent approaches can be found in
the surveys \cite{Lang08,Rizzo06}. Further  subdivision can be
done into the considerations  accounting for any type correlations
between the parameters of the low energy effective interactions
and that of assuming complete independence of them. Because of a
large amount of fitting parameters the latter are less
predictable.

Now, for a convenience of readers let us present the results of
fits of the $Z'$ parameters  in terms of the popular notations
\cite{Leike,Lang08}. The Lagrangian reads
\begin{eqnarray}\label{avstandard}
{\cal L}_{Z\bar{f}f}&=&\frac{1}{2} Z_\mu\bar{f}\gamma^\mu\left[
(v^{\mathrm{SM}}_f+ \Delta^V_f) - \gamma^5
(a^{\mathrm{SM}}_f+\Delta^A_f) \right]f, \nonumber\\
{\cal L}_{Z'\bar{f}f}&=&\frac{1}{2} Z'_\mu\bar{f}\gamma^\mu\left[
(v'_f-\gamma^5 a'_f)\right]f,
\end{eqnarray}
with the SM values of the $Z$ couplings
\[
v^{\mathrm{SM}}_f = \frac{e\left(T_{3f}-2Q_f
\sin^2\theta_W\right)}{\sin\theta_W\cos\theta_W},\qquad
a^{\mathrm{SM}}_f = \frac{e\,T_{3f}}{\sin\theta_W\cos\theta_W},
\]
where $e$ is the positron charge, $Q_f$ is the fermion charge in
the units of $e$, $T_{3f}=1/2$ for the neutrinos and $u$-type
quarks, and $T_{3f}=-1/2$ for the charged leptons and $d$-type
quarks.

\begin{figure}
\centering
  \includegraphics[bb= 0 0 240 185,width=.4\textwidth]{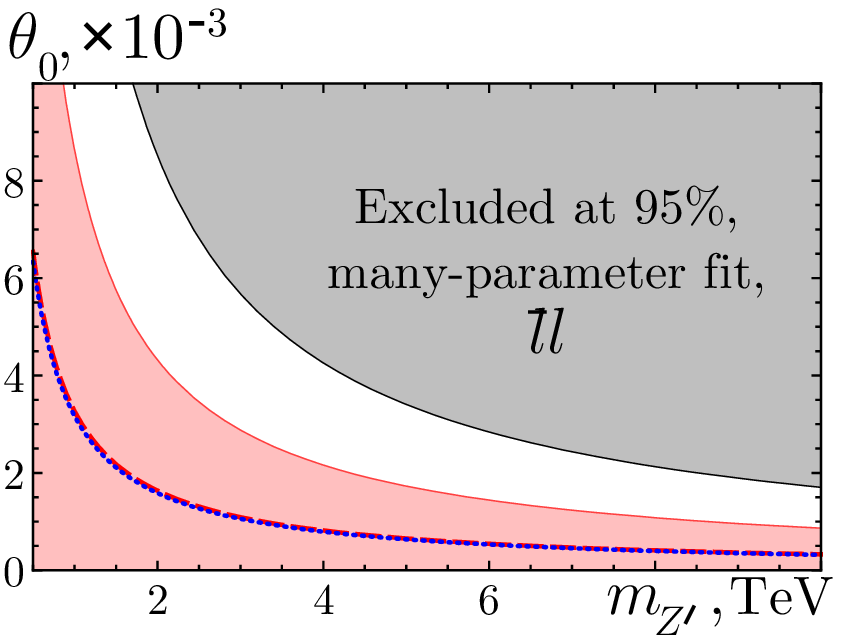}
  \includegraphics[bb= 0 0 240
  185,width=.4\textwidth]{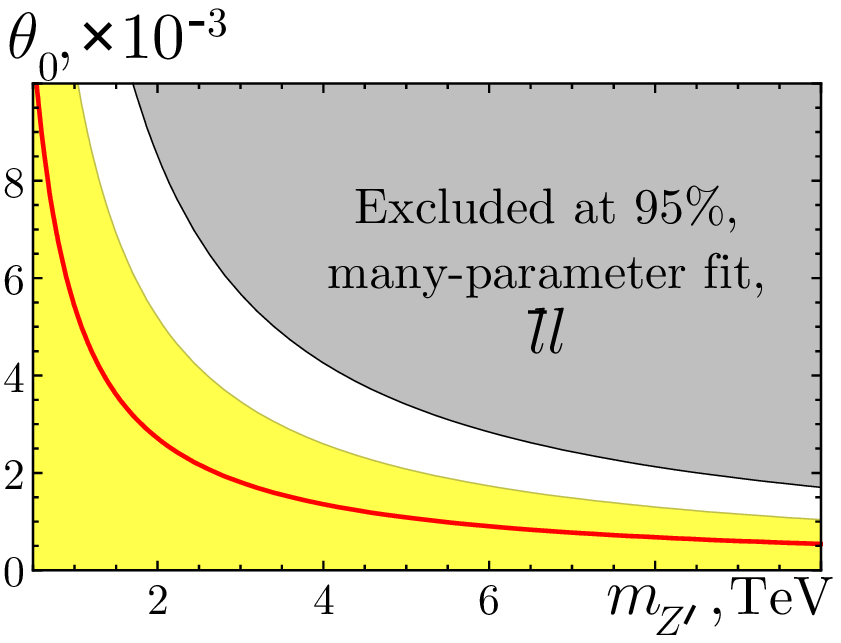}\\
  \includegraphics[bb= 0 0 240 185,width=.4\textwidth]{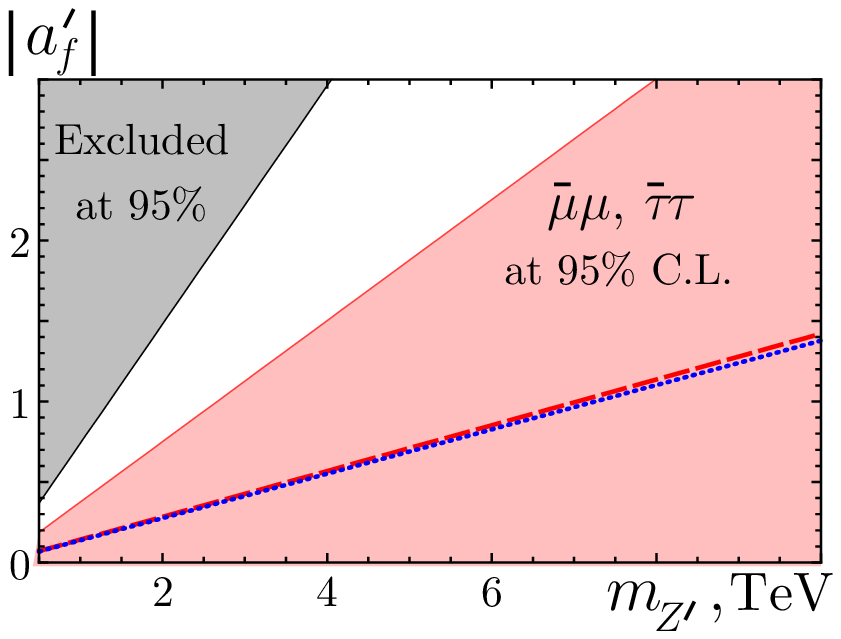}
  \includegraphics[bb= 0 0 240 185,width=.4\textwidth]{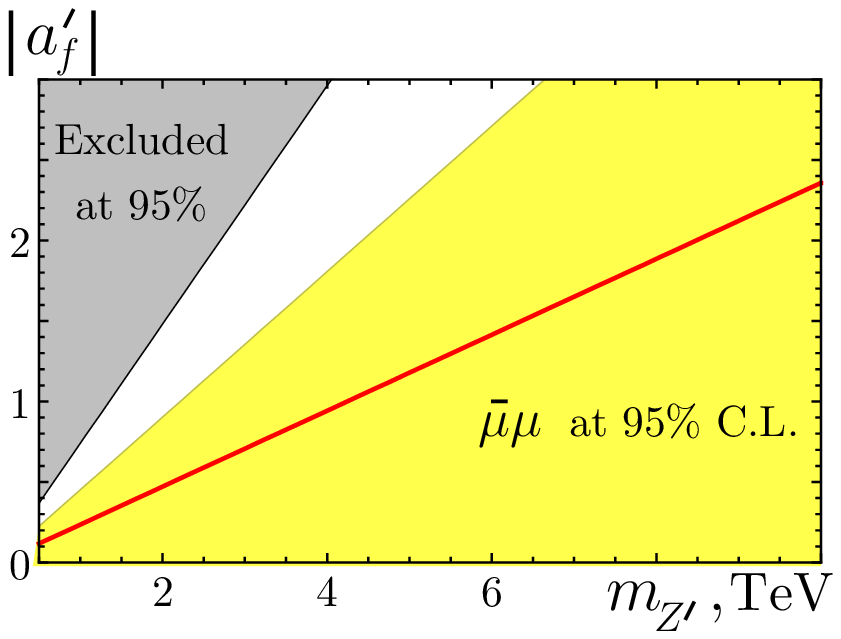}\\
  \includegraphics[bb= 0 0 240 185,width=.4\textwidth]{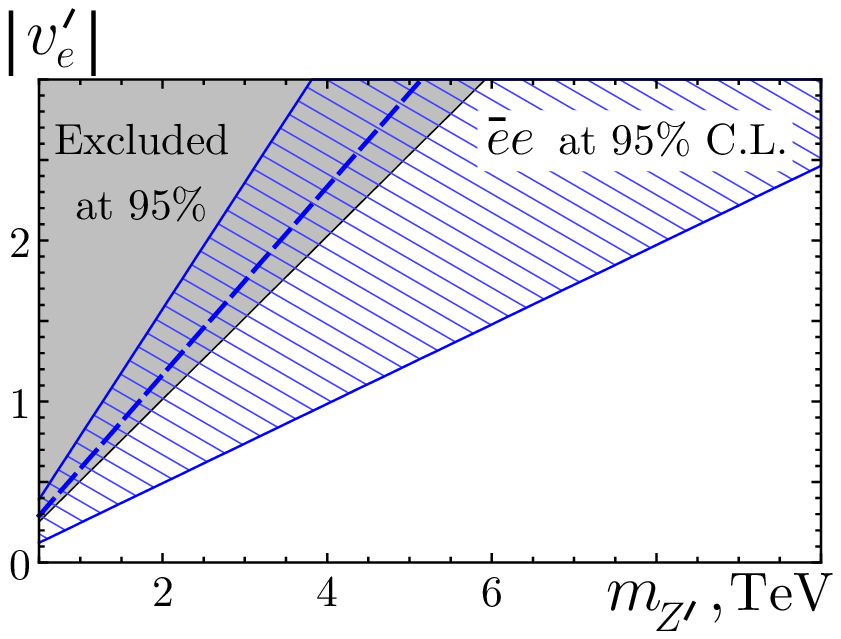}
  \includegraphics[bb= 0 0 240 185,width=.4\textwidth]{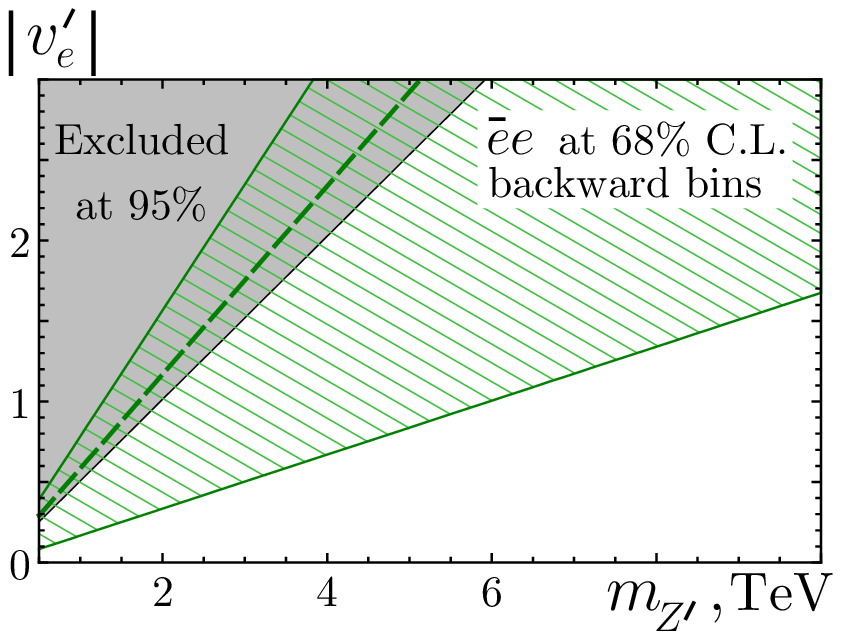}
  \caption{The maximum likelihood values and the confidence
intervals for the $Z-Z'$ mixing angle ($\theta_0$), the vector
coupling to electron current ($v'_e$), and the axial-vector
couplings to the SM fermions ($a'_f$) by the LEP 2 data. The
values excluded at 95\% CL by the many-parameter fit of $e^+e^-\to
l^+l^-$ are shown in gray. The results of fits based on the
one-parameter observables are shown in pink for $e^+e^-\to
\mu^+\mu^-,\tau^+\tau^-$ (with the maximum likelihood value as the
dashed red line), in yellow for $e^+e^-\to \mu^+\mu^-$  (with the
maximum likelihood value as the solid red line), and in blue for
$e^+e^-\to e^+e^-$  (with the maximum likelihood value as the
dashed blue line). The maximum likelihood values and the 1$\sigma$
CL area for the many-parameter fit of backward bins of $e^+e^-\to
e^+e^-$ are also shown in green. The blue dotted lines correspond
to the maximum likelihood values obtained from the LEP1 data.
}\label{fig:fitFigs}
\end{figure}

\begin{table}
  \centering
  \caption{The summary of the fits of the LEP data for
  the maximum likelihood values of the $Z'$ couplings (\ref{avstandard}) to the SM fermions
  and of the $Z-Z'$ mixing angle $\theta_0$.
  $M=\frac{m_{Z'}}{1\, \mathrm{TeV}}$ denotes the unknown value of the $Z'$ mass in TeV units.}\label{fitMLV}
\begin{tabular}{|l|c|c|c|c|}
  \hline
  Data & $|\theta_0|$, $\times 10^{-3}$ & $|v'_e|$, $\times 10^{-1}$ & $|a'_f|$, $\times 10^{-1}$ & $\Delta^A_e$, $\times 10^{-3}$ \\
  \hline\hline
  \multicolumn{5}{|c|}{LEP1} \\
  \hline
  $e^-e^+$                  & ${3.17}{M}^{-1}$ & -     & $1.38M$ & 0.437 \\
  \hline
  \multicolumn{5}{|c|}{LEP2, one-parameter fits} \\
  \hline
  $e^-e^+$                  & -              & $5.83M$ & -        & - \\
  $\mu^-\mu^+$              & $5.42{M}^{-1}$ & -       & $2.36M$ & 1.278 \\
  $\mu^-\mu^+,\tau^-\tau^+$ & $3.27{M}^{-1}$ & -       & $1.42M$ & 0.464 \\
  \hline
  \multicolumn{5}{|c|}{LEP2, many-parameter fits} \\
  \hline
  $e^-e^+$, $z<0$           & -              & $5.84M$ & -        & - \\
  \hline
\end{tabular}
\end{table}
\begin{table}
  \centering
  \caption{The summary of the fits of the LEP data for
  the confidence intervals for the $Z'$ couplings (\ref{avstandard}) to the SM fermions
  and for the $Z-Z'$ mixing angle $\theta_0$.
  $M=\frac{m_{Z'}}{1\, \mathrm{TeV}}$ denotes the unknown value of the $Z'$ mass in TeV units.}\label{fitCLI}
\begin{tabular}{|l|c|c|c|c|c|}
  \hline
  Data & CL & $|\theta_0|$, $\times 10^{-3}$ & $|v'_e|$, $\times 10^{-1}$ & $|a'_f|$, $\times 10^{-1}$ & $\Delta^A_e$, $\times 10^{-3}$ \\
  \hline\hline  \multicolumn{6}{|c|}{LEP1} \\
  \hline  $e^-e^+$ & 68\% & $(0;4.48){M}^{-1}$ & - & $(0;1.95)M$ & $(0;0.873)$ \\
  \hline  \multicolumn{6}{|c|}{LEP2, one-parameter fits} \\
  \hline  $e^-e^+$ & 95\% & - & $(2.46;7.87)M$ & - & - \\
  \hline $\mu^-\mu^+$ & 95\% & $(0;10.39){M}^{-1}$ & - & $(0;4.52)M$ & $(0;4.694)$ \\
  \hline $\mu^-\mu^+$,&&&&& \\
  $\tau^-\tau^+$ & 95\% & $(0;8.64){M}^{-1}$ & - & $(0;3.75)M$ & $(0;3.244)$ \\
  \hline  \multicolumn{6}{|c|}{LEP2, many-parameter fits} \\
  \hline  $e^-e^+$, &&&&& \\
  $\mu^-\mu^+$, & 95\% & $(0;17.03){M}^{-1}$ & $(0;5.06)M$ & $(0;7.40)M$ & $(0;12.607)$ \\
  $\tau^-\tau^+$ &&&&& \\
  \hline $e^-e^+$, &&&&& \\
  $z<0$ & 68\% & $(0;27.61){M}^{-1}$ & $(1.68; 7.83)M$ & $(0;12.00)M$ & $(0;33.1288)$ \\
  \hline
\end{tabular}
\end{table}
The results of the fits of the $Z'$ couplings to the SM leptons
obtained from the analysis of the LEP experiments are adduced in
the Tables \ref{fitMLV}-\ref{fitCLI} and Fig. \ref{fig:fitFigs}.
Remind that due to the universality of the axial-vector coupling
$a_f$ the same estimates take also place for quarks. First of all,
one parameter fits of LEP experiments as well as the
many-parameter fit for the $e^+e^-$ backward bins show the hints
of the $Z'$ boson at the 1-2$\sigma$ CL. Due to this fact, the
fits allow to determine the maximum likelihood values of $Z'$
parameters. In spite of uncertainties, these values can be used as
a guiding line for the estimation of possible $Z'$ effects in the
LHC experiments. The maximum likelihood values are given in Table
\ref{fitMLV}. As is seen, different fits and processes lead to the
comparable values of the $Z'$ parameters.

In Table \ref{fitCLI} we present the confidence intervals for the
fitted parameters. With this Table one is able to estimate the
uncertainty of the $Z'$ couplings as well as the lower bounds on
the parameters.

In Fig. \ref{fig:fitFigs} the maximum likelihood values and the CL
intervals are drawn for the different values of the $Z'$ mass. All
the plots exploit the same color scheme. The values excluded at
95\% CL by the many-parameter fit of all the LEP2 leptonic
processes $e^+e^-\to l^+l^-$ are shown in gray. The 95\%
confidence intervals from the one-parameter fit of LEP2
$e^+e^-\to\mu^+\mu^-,\tau^+\tau^-$ are drawn in pink with the
maximum likelihood values as the dashed red line. The
corresponding results with taking into account the $\mu^+\mu^-$
process only are shown in yellow with solid red line for the
maximum likelihood values. The maximum likelihood values from the
LEP1 experiments are represented as dotted blue line. The 95\%
confidence interval from the one-parameter fit of the LEP2 Bhabha
scattering is shown as the blue crosshatched region with the
maximum likelihood values as the dashed blue line. Finally, the
68\% confidence interval and the maximum likelihood values from
the many-parameter fit of the backward bins of LEP2 Bhabha
scattering are shown in green.

Now we compare the above results with the ones of other fits
accounting for the $Z'$ presence. As it was mentioned in
Introduction, LEP collaborations have determined the model
dependent low bounds on the $Z'$ mass which vary in a wide energy
interval dependently on  a model. The same has also been done for
Tevatron experiments. The  modern low bound is $m_{Z'} \ge 850$
GeV. It is also well known that though almost all the present day
data are described by the SM \cite{EWWG,OPAL,DELPHI,LEP1}, the
overall fit to the standard model is not very good. In Ref.
\cite{Ferroglia} it was showed that the large difference in
$\sin^2 \theta^\mathrm{lept}_\mathrm{eff}$ from the
forward-backward asymmetry $A^b_{fb}$ of the bottom quarks and the
measurements from the SLAC SLD experiment can be explained for
physically reasonable Higgs boson mass if one allows for one or
more extra $U(1)$ fields, that is $Z'$. A specific model to
describe $Z'$ physics of interest was proposed which introduces
two type couplings to the hyper charge $Y$ and to the
baryon-minus-lepton number $B-L$. Within this model by using a
number of precision measurements from LEP1, LEP2, SLD and Tevatron
experiments the parameters $a_Y$ and $a_{B-L}$ of the model were
fitted. The presence of $Z'$ was not excluded at 68\% CL. The
value  of  $a_Y$ was estimated to be of the same order of
magnitude as in  our analysis and is comparable with values of
other parameters detected in the LEP experiments.  The erroneous
claim that $a_Y$ is two order less then the value derived from our
Table 4 is, probably,  a consequence of some missed factors. The
upper limit on the mass was also obtained $m_{Z'} \le 2.6 $ TeV .

These two analyzes are different but complementary.  A common
feature of them is an accounting for the $Z'$ gauge boson as a
necessary element of the data fits. The results are in accordance
at 68-95\% CL with the existence of the not heavy $Z'$ which has a
good chance to be discovered at Tevatron and/or LHC.





\section*{Appendix. RG relations in a theory with different mass
scales} \addcontentsline{toc}{section}{Appendix. RG relations in a
theory with different mass scales}

In this Appendix we are going to investigate the Yukawa model with
a heavy scalar field $\chi$ and a light scalar field $\varphi$
\cite{YAF2001}. The goals of our investigation are two fold: 1) to
derive the one-loop RG relation for the four-fermion scattering
amplitude in the decoupling region and 2) to find out the
possibility of reducing this relation in the equation for vertex
describing the scattering of light particles on the external field
when the mixing between heavy and light virtual states takes
place.

The Lagrangian of the model reads
\begin{eqnarray}\label{app1}
{\cal L}&=&\frac{1}{2}{\left( \partial_{\mu}\varphi \right) }^{2}-
\frac{m^{2}}{2}{\varphi}^{2}-\lambda{\varphi}^{4}+
\frac{1}{2}{\left(
\partial_{\mu}\chi \right) }^{2}- \frac{{\Lambda}^{2}}{2}{\chi}^{2}-
\xi{\chi}^{4}\nonumber \\ &&+\rho
{\varphi}^{2}{\chi}^{2}+{\bar\psi}\left(
i\partial_{\mu}\gamma_{\mu}-M-G_{\varphi}\varphi- G_{\chi}\chi
\right) \psi,
\end{eqnarray}
where $\psi$ is a Dirac spinor field, and $\Lambda\gg m,M$.

\begin{figure}
\begin{center}
\includegraphics[bb= 0 0 600 600,width=.2\textwidth]
{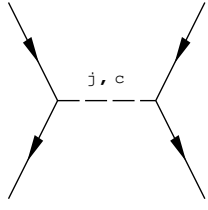}
  \caption{Tree level contribution to the four-fermion amplitude.}
\label{fig:tree}
\end{center}
\end{figure}

\begin{figure}
\begin{center}
\includegraphics[bb= 0 0 1800 600,width=.46372\textwidth]
{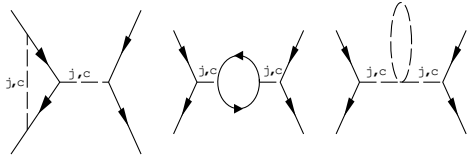}
  \caption{The one-loop level contribution to the one-particle
           reducible four-fermion amplitude.}
\label{fig:loop}
\end{center}
\end{figure}

Consider the four-fermion scattering
$\bar\psi\psi\to\bar\psi\psi$. The $S$-matrix element at the
one-loop level is given by
\begin{eqnarray}\label{app2}
{\hat S}&=&-\frac{i}{2}\int\frac{dp_{1}}{{\left( 2\pi\right)
}^{4}}\ldots\frac{dp_{4}}{{\left( 2\pi\right) }^{4}}{\left(
2\pi\right) }^{4}\delta\left( p_{1}+...+p_{4}\right)
\left[ S_\mathrm{1PR}+ S_\mathrm{box}\right],\nonumber \\
S_\mathrm{1PR}&=&\sum\limits_{{\phi}_{1},{\phi}_{2}=\varphi,
\chi}G_{{\phi}_{1}} G_{{\phi}_{2}}\left(
\frac{{\delta}_{{\phi}_{1}{\phi}_{2}}}{s-m^2_{{\phi}_{1}}}
+\frac{\Pi_{\phi_1\phi_2}(s)}{(s- m^2_{\phi_1})(s-
m^2_{\phi_2})}\right)
 \nonumber\\&&
\times {\bar\psi}\left(p_{4}\right)\left[1+\Gamma\left(
p_{3}, -p_{4}- p_{3}\right)  \right] \psi\left(p_{3}\right)\nonumber \\
&&\times {\bar\psi}\left( p_{1}\right) \left[1+\Gamma\left( p_{2},
-p_{1}- p_{2}\right)  \right] \psi\left( p_{2}\right),
\end{eqnarray}
where $s={\left( p_{1}+p_{2}\right) }^{2}$, $S_\mathrm{1PR}$ is
the contribution from the one-particle reducible diagrams shown in
Figs. \ref{fig:tree}-\ref{fig:loop}, and $S_\mathrm{box}$ is the
contribution from box diagrams.
The one-loop polarization operator of scalar fields
${\Pi}_{{\phi}_{1}{\phi}_{2}}$ and the one-loop vertex function
$\Gamma$ are related to the Green functions as
\begin{eqnarray}\label{app3}
D_{{\phi}_{1}{\phi}_{2}}\left( s \right)
&=&\frac{{\delta}_{{\phi}_{1}{\phi}_{2}}}{s-m^2_{
{\phi}_{1}}}+\frac{1}{s-
m^2_{{\phi}_{1}}}{\Pi}_{{\phi}_{1}{\phi}_{2}}\left( s \right)
\frac{1}{s- m^2_{{\phi}_{2}}},\nonumber \\ G_{\phi\phi\psi}\left(
p,q \right)&=&-\sum\limits_{{\phi}_{1} }G_{{\phi}_{1}}
D_{{\phi}_{1}\phi}\left( q^{2} \right) S_{\psi}\left( p \right)
\left(1+\Gamma\left( p, q\right)  \right) S_{\psi} \left(
-p-q\right),
\end{eqnarray}
where $ S_{\psi}$ is the spinor propagator in the momentum
representation.

\subsection*{Renormalization of the model}

The renormalized fields, masses and charges are defined as follows
\begin{eqnarray}\label{app4}
&&\left( \begin{array}{c}\varphi \\ \chi
\end{array}\right)= Z_\phi^{-1/2}\left( \begin{array}{c}\varphi_0 \\
\chi_0 \end{array}\right),\qquad \left(
\begin{array}{c}G_\varphi
\\ G_\chi
\end{array}\right)= Z_G^{-1}\left( \begin{array}{c}G_{\varphi, 0} \\
G_{\chi, 0} \end{array}\right),
 \nonumber\\&&
\psi=Z_\psi^{-1/2}\psi_0,\qquad M^2= M_0^2-\delta M^2,
 \nonumber\\&&
m^2=m_0^2-\delta m^2,\qquad\Lambda^2=\Lambda_0^2-\delta\Lambda^2,
\end{eqnarray}
where subscript 0 marks the corresponding bare quantities.

Using the dimensional regularization (the dimension of the
momentum space is $D=4-\varepsilon$) and the
$\overline{\mathrm{MS}}$ renormalization scheme one can compute
the renormalization constants
\begin{eqnarray}\label{app5}
Z_\psi&=&1-\frac{1}{16\pi^2\varepsilon}\left(G_\varphi^2+G_\chi^2\right),
\nonumber\\ \delta
M^2&=&\frac{3}{8\pi^2\varepsilon}\left(G_\varphi^2+G_\chi^2\right)M^{2}, \nonumber\\
Z_\phi^{1/2}&=&1-\frac{1}{8\pi^2\varepsilon}\left(\begin{array}{cc}
G_\varphi^2& 2G_\varphi
G_\chi\frac{\Lambda^2-6M^2}{\Lambda^2-m^2}\\
-2G_\varphi G_\chi\frac{m^2-6M^2}{\Lambda^2-m^2}&
G_\chi^2\end{array}\right), \nonumber\\ \delta
m^2&=&\frac{1}{4\pi^2\varepsilon}
\left[\left(G_\varphi^2+6\lambda\right)m^2-6G_\varphi^2M^2-\rho\Lambda^2\right],
\nonumber\\
\delta\Lambda^2&=&\frac{1}{4\pi^2\varepsilon}
\left[\left(G_\chi^2+6\xi\right)\Lambda^2-6G_\chi^2M^2-\rho
m^2\right], \nonumber\\
Z_G^{-1}&=&\left[1-\frac{3}{16\pi^2\varepsilon}\left(G_\varphi^2+G_\chi^2\right)\right]
{\left(Z_\phi^{1/2}\right)}^\mathrm{T}.
\end{eqnarray}

From Eq. (\ref{app5}) we obtain the appropriate $\beta$ and
$\gamma$ functions at the one-loop level:
\begin{eqnarray}\label{app6}
\beta_\varphi&=&\frac{dG_\varphi}{d\log\kappa}=
\frac{G_\varphi}{16\pi^2}\left(5G_\varphi^2+3G_\chi^2-4\frac{m^2-6M^2}{\Lambda^2-
m^2}G_\chi^2\right),
\nonumber\\
\beta_\chi&=&\frac{dG_\chi}{d\log\kappa}=
\frac{G_\chi}{16\pi^2}\left(5G_\chi^2+3G_\varphi^2+4\frac{\Lambda^2-6M^2}{\Lambda^2-m^2}G_\varphi^2\right),
\nonumber\\
\gamma_m&=&-\frac{d\log m^2}{d\log\kappa}=
-\frac{1}{4\pi^2}\left(G_\varphi^2\frac{m^2-6M^2}{m^2}+6\lambda-\rho\frac{\Lambda^2}{m^2}\right),
\nonumber\\
\gamma_\Lambda&=&-\frac{d\log\Lambda^2}{d\log\kappa}=
-\frac{1}{4\pi^2}\left(G_\chi^2\frac{\Lambda^2-6M^2}{\Lambda^2}+6\xi-\rho\frac{m^2}{\Lambda^2}\right),
\nonumber\\
\gamma_\psi&=&-\frac{d\log\psi}{d\log\kappa}=
\frac{1}{32\pi^2}\left(G_\varphi^2+G_\chi^2\right).
\end{eqnarray}
Then, the $S$-matrix element can be expressed in terms of the
renormalized quantities. The contribution from the one-particle
reducible diagrams becomes
\begin{eqnarray}\label{app7}
S_\mathrm{1PR}&=&\sum\limits_{\phi_1,\phi_2}G_{\phi_1}G_{\phi_2}
\left(\frac{\delta_{\phi_1\phi_2}}{s-m^2_{\phi_1}}
+\frac{\Pi_{\phi_1\phi_2}^\mathrm{(fin)}(s)}{(s-m^2_{\phi_1})(s-m^2_{\phi_2})}\right)
\nonumber\\&&\times
\bar\psi(p_4)\left[1+\Gamma^\mathrm{(fin)}\left(p_3,-p_4-p_3\right)\right]\psi(p_3)
\nonumber\\&&\times
\bar\psi(p_1)\left[1+\Gamma^\mathrm{(fin)}\left(p_2,-p_1-p_2\right)\right]\psi(p_2),
\end{eqnarray}
where the functions $\Pi_{\phi_1\phi_2}^\mathrm{(fin)}$ and
$\Gamma^\mathrm{(fin)}$ are the expressions $\Pi_{\phi_1\phi_2}$
and $\Gamma$ without the terms proportional to $1/\varepsilon$.
Since the quantity $S_\mathrm{box}$ is finite, the renormalization
leaves it without changes.

Introducing the RG operator at the one-loop level \cite{bando93}
\begin{eqnarray}\label{app8}
{\cal
D}&=&\frac{d}{d\log\kappa}=\frac{\partial}{\partial\log\kappa}+
{\cal D}^{(1)} =\frac{\partial}{\partial\log\kappa} \nonumber\\&&
+\sum\limits_\phi\beta_\phi\frac{\partial}{\partial G_\phi}
-\gamma_m\frac{\partial}{\partial\log m^2}
-\gamma_\Lambda\frac{\partial}{\partial\log\Lambda^2}
-\gamma_\psi\frac{\partial}{\partial\log\psi}
\end{eqnarray}
we determine that the following relation holds for the $S$-matrix
element
\begin{equation}\label{app9}
{\cal D}\left( S_\mathrm{1PR}+S_\mathrm{box}\right)
=\frac{\partial S_\mathrm{1PR}^{(1)}}{\partial\log\kappa}+{\cal
D}^{(1)} S_\mathrm{1PR}^{(0)}=0,
\end{equation}
where the $S_\mathrm{1PR}^{(0)}$ and the $S_\mathrm{1PR}^{(1)}$
are the contributions to the $S_\mathrm{1PR}$ at the tree level
and at the one-loop level, respectively:
\begin{equation}\label{app10}
S_\mathrm{1PR}^{(0)}=
\left(\frac{G_\varphi^2}{s-m^2}+\frac{G_\chi^2}{s-\Lambda^2}\right)
\bar\psi\psi\times\bar\psi\psi,
\end{equation}
\begin{eqnarray}\label{app11}
\frac{\partial S_\mathrm{1PR}^{(1)}}{\partial\log\kappa}&=&
\frac{\bar\psi\psi\times\bar\psi\psi}{4\pi^2}
\left[ -\left(G_\varphi^2+G_\chi^2\right)
\left(\frac{G_\varphi^2}{s-m^2}+\frac{G_\chi^2}{s-\Lambda^2}\right)
\right.\nonumber\\&&
+\frac{G_\varphi^2\left(\rho\Lambda^2-6\lambda
m^2+G_\varphi^2\left(6M^2-s\right)\right)}{\left(s-m^2\right)^2}
\nonumber\\&&
+\frac{2G_\varphi^2G_\chi^2\left(6M^2-s\right)}{\left(s-m^2\right)\left(s-\Lambda^2\right)}
\nonumber\\&&\left. +\frac{G_\chi^2\left(\rho
m^2-6\xi\Lambda^2+G_\chi^2\left(6M^2-s\right)\right)}{\left(s-\Lambda^2\right)^2}\right].
\end{eqnarray}

The first term in Eq. (\ref{app11}) is originated from the
one-loop correction to the fermion-scalar vertex. The rest terms
are connected with the polarization operator of scalars. The third
term describes the one-loop mixing between the scalar fields. It
is canceled in the RG relation (\ref{app9}) by the mass-dependent
terms in the $\beta$ functions produced by the non-diagonal
elements in $Z_\phi$.

Eq. (\ref{app9}) is the consequence of the renormalizability of
the model. It insures the leading logarithm terms of the one-loop
$S$-matrix element to reproduce the appropriate tree-level
structure. In contrast to the familiar treatment we are not going
to improve scattering amplitudes by solving Eq. (\ref{app9}). We
will use it as an algebraic identity implemented in the
renormalizable theory. Naturally if one knows the explicit
couplings expressed in terms of the basic set of parameters of the
model, this RG relation is trivially fulfilled. But the situation
changes when the couplings are represented by unknown arbitrary
parameters. In this case the RG relations are the algebraic
equations dependent on these parameters and appropriate $\beta$
and $\gamma$ functions. In the presence of a symmetry the number
of $\beta$ and $\gamma$ functions is less than the number of RG
relations. So, one has non trivial system of equations relating
the unknown couplings. For example, such a scenario is realized
for the gauge coupling. Although the considered simple model has
no gauge couplings, we are able to demonstrate the general
procedure of deriving the RG relations.

\subsection*{Decoupling of the heavy field}

At energies $s\ll\Lambda^2$ the heavy scalar field $\chi$ is
decoupled. This means, that the four-fermion scattering amplitude
is described by the model with no heavy field $\chi$ plus terms of
the order $s/\Lambda^2$. At the tree level, this is the obvious
consequence of the expansion of the heavy scalar propagator
\begin{equation}\label{app12}
\frac{1}{s-\Lambda^2}\to
-\frac{1}{\Lambda^2}\left[1+O\left(\frac{s}{\Lambda^2}\right)\right],
\end{equation}
which is resulted in the effective contact four-fermion
interaction in Eq. (\ref{app10})
\begin{equation}\label{app13}
{\cal L}_\mathrm{eff}=-\alpha\,\bar\psi\psi\times\bar\psi\psi,
\quad\alpha=\frac{G_\chi^2}{\Lambda^2}.
\end{equation}
So, the tree level contribution to the scattering amplitude
becomes
\begin{equation}\label{app14}
S_\mathrm{1PR}^{(0)}= \left[
\frac{G_\varphi^2}{s-m^2}-\alpha+O\left(\frac{s}{\Lambda^4}\right)
\right] \bar\psi\psi\times\bar\psi\psi,
\end{equation}
and the lowest order effects of the heavy scalar in the decoupling
region are described by the parameter $\alpha$, only.

Decoupling of heavy particles is present also at the level of
radiative corrections. The radiative corrections are generally
described by various loop integrals in the momentum space (the
Passarino--Veltman functions). Considering a Passarino--Veltman
function with at least one heavy mass $Lambda$ inside loop in the
low-energy limit, one can see the following asymptotic behavior:
the function splits into 1) possible energy-independent divergent
part (including also $\log\Lambda$) and 2) energy-dependent finite
part which can be expanded by inverse powers of $\Lambda$ and
vanishes at the small energies. The important property is that the
$\log\Lambda$-term in the divergent part reproduces the logarithm
of the cut-off scale. So, such a potentially large term has to be
automatically absorbed by the renormalization at low energies and
leads to no observable effects. However, if the renormalization is
actually performed at high energies (as in the
$\overline{\mathrm{MS}}$ renormalization scheme) the potentially
large $\log\Lambda$-terms should be re-summed manually by the
redefinition of the physical couplings and masses at the scale
$\Lambda$.

What is the form of the RG relations in the limit of large
$\Lambda$? The method of constructing the RG equation in the
decoupling region was proposed in \cite{bando93}. It introduces
the redefinition of the parameters of the model allowing to remove
all the heavy particle loop contributions to Eq. (\ref{app11}).
Let us define a new set of fields, charges and masses
$\tilde\psi$, $\tilde{G}_\varphi$, $\tilde{G}_\chi$,
$\tilde\Lambda$, $\tilde{m}$, $\tilde{M}$
\begin{eqnarray}\label{app15}
G_\varphi^2&=&\tilde{G}_\varphi^2
\left(1+\frac{3\tilde{G}_\chi^2}{16\pi^2}
\log\frac{\kappa^2}{\tilde\Lambda^2}+\ldots\right), \nonumber\\
G_\chi^2&=&\tilde{G}_\chi^2
\left(1+\frac{3\tilde{G}_\chi^2}{16\pi^2}
\log\frac{\kappa^2}{\tilde\Lambda^2}+\ldots\right), \nonumber\\
m^2&=&\tilde{m}^2\left(1-\frac{\tilde\rho}{8\pi^2}\frac{\tilde\Lambda^2}{\tilde{m}^2}
\log\frac{\kappa^2}{\tilde\Lambda^2}+\ldots\right), \nonumber\\
\Lambda^2&=&\tilde\Lambda^2 \left(1+\frac{3\tilde\xi}{4\pi^2}
\log\frac{\kappa^2}{\tilde\Lambda^2}+\ldots\right), \nonumber\\
\psi&=&\tilde\psi\left(1-\frac{\tilde{G}_\chi^2}{64\pi^2}
\log\frac{\kappa^2}{\tilde\Lambda^2}+\ldots\right),
\end{eqnarray}
where dots stand for the higher powers of $\log\Lambda$
responsible for the decoupling at higher loop orders.

The differential operator (\ref{app8}) ban be rewritten in terms
of these new low-energy parameters:
\begin{eqnarray}\label{app16}
{\cal D}&=&\frac{\partial}{\partial\log\kappa}+{\tilde{\cal
D}}^{(1)}= \frac{\partial}{\partial\log\kappa}
+\sum\limits_\phi\tilde\beta_\phi\frac{\partial}{\partial\tilde{G}_\phi}
\nonumber\\&&
-\tilde\gamma_m\frac{\partial}{\partial\log\tilde{m}^2}
-\tilde\gamma_\Lambda\frac{\partial}{\partial\log\tilde\Lambda^2}
-\tilde\gamma_\psi\frac{\partial}{\partial\log\tilde\psi}
\end{eqnarray}
where $\tilde\beta$ and $\tilde\gamma$ functions are obtained from
the one-loop relations (\ref{app6}) and (\ref{app15})
\begin{eqnarray}\label{app17}
&& \tilde\beta_\varphi=\frac{1}{16\pi^2}
\left(5\tilde{G}_\varphi^{3}-4\frac{\tilde{m}^2-6\tilde{M}^2}{\tilde\Lambda^2-\tilde{m}^2}\tilde{G}_\varphi\tilde{G}_\chi^2\right),
\nonumber\\
&& \tilde\beta_\chi=\frac{1}{16\pi^2}
\left(2\tilde{G}_\chi^3+\left(3+4\frac{\tilde\Lambda^2-6\tilde{M}^2}{\tilde\Lambda^2-\tilde{m}^2}\right)
\tilde{G}_\chi\tilde{G}_\varphi^2\right),
\nonumber\\
&& \tilde\gamma_m=-\frac{1}{4\pi^2}
\left(\tilde{G}_\varphi^2\frac{\tilde{m}^2-6\tilde{M}^2}{\tilde{m}^2}+6\tilde\lambda\right),
\nonumber\\
&& \tilde\gamma_\Lambda=-\frac{1}{4\pi^2}
\left(\tilde{G}_\chi^2\left(1-6\frac{\tilde{M}^2}{\tilde\Lambda^2}\right)
-\tilde\rho\frac{\tilde{m}^2}{\tilde\Lambda^2}\right),
\nonumber\\
&& \tilde\gamma_\psi=\frac{1}{32\pi^2}\tilde{G}_\varphi^2.
\end{eqnarray}
Hence, one immediately notices that $\tilde\beta$ and
$\tilde\gamma$ functions contain only the light particle loop
contributions, and all the heavy particle loop terms are
completely removed from them. The $S$-matrix element expressed in
terms of new parameters satisfies the following RG relation
\begin{equation}\label{app18}
{\cal D}\left(S_\mathrm{1PR}+S_\mathrm{box}\right)
=\frac{\partial\tilde{S}_\mathrm{1PR}^{(1)}}{\partial\log\kappa}
+{\tilde{\cal D}}^{(1)}\tilde{S}_\mathrm{1PR}^{(0)}=0,
\end{equation}
\begin{equation}\label{app19}
\tilde{S}_\mathrm{1PR}^{(0)}=
\left(\frac{\tilde{G}_\varphi^2}{s-\tilde{m}^2} -\tilde\alpha
+O\left(\frac{s^2}{\tilde\Lambda^4}\right)\right)
{\bar{\tilde\psi}}\tilde\psi\times{\bar{\tilde\psi}}\tilde\psi,
\end{equation}
\begin{eqnarray}\label{app20}
\frac{\partial\tilde{S}_\mathrm{1PR}^{(1)}}{\partial\log\kappa}&=&
\frac{{\bar{\tilde\psi}}\tilde\psi\times{\bar{\tilde\psi}}\tilde\psi}{4\pi^2}
\left(-\frac{\tilde{G}_\varphi^4}{s-\tilde{m}^2}
\right.\nonumber\\&&\left.
+\frac{\tilde{G}_\varphi^2\left(-6\tilde\lambda\tilde{m}^2
+\tilde{G}_\varphi^2\left(6\tilde{M}^2-s\right)\right)}{\left(s-\tilde{m}^2\right)^2}
+\tilde\alpha\tilde{G}_\varphi^2 \right.\nonumber\\&&\left.
-\frac{2\tilde{G}_\varphi^2\tilde\alpha\left(6\tilde{M}^2-s\right)}{s-\tilde{m}^2}
+O\left(\frac{s^2}{\tilde\Lambda^4}\right)\right),
\end{eqnarray}
where $\tilde\alpha=\tilde{G}_\chi^2/\tilde\Lambda^2$ is the
redefined effective four-fermion coupling. As one can see, Eq.
(\ref{app20}) includes all the terms of Eq. (\ref{app11}) except
for the heavy particle loop contributions. It depends on the low
energy quantities $\tilde\psi$, $\tilde{G}_\varphi$,
$\tilde\alpha$, $\tilde\lambda$, $\tilde{m}$, $\tilde{M}$. The
first and the second terms in Eq. (\ref{app20}) are just the
one-loop amplitude calculated within the model with no heavy
particles. The third and the fourth terms describe the light
particle loop correction to the effective four-fermion coupling
and the mixing of heavy and light virtual fields.

\subsection*{Elimination of the one-loop scalar field mixing}

Due to the mixing term it is impossible to split the RG relation
(\ref{app18}) for the $S$-matrix element into the one for
vertices. Hence, we are not able to consider Eq. (\ref{app18}) in
the framework of the scattering of light particles on an external
field induced by the heavy virtual scalar. But this is an
important step in deriving the RG relation for EL parameters.
Fortunately, there is a simple procedure allowing to avoid the
mixing in Eq. (\ref{app20}). The way is to incorporate the
diagonalization of the leading logarithm terms of the scalar
polarization operator into the redefinition of the
$\tilde\varphi$, $\tilde\chi$, $\tilde{G}_\varphi$,
$\tilde{G}_\chi$:
\begin{eqnarray}\label{app21}
\left( \begin{array}{c}\varphi \\ \chi \end{array}\right)&=&
{\zeta}^{1/2}\left( \begin{array}{c}{\tilde\varphi} \\
{\tilde\chi}
\end{array}\right),\nonumber\\ \left( \begin{array}{c}G_\varphi \\ G_\chi
\end{array}\right)&=&\left[1+\frac{3\tilde{G}_\chi^2}{32\pi^2}
\log\frac{\kappa^2}{\tilde\Lambda^2}\right]
\left(\zeta^{-1/2}\right)^\mathrm{T}
\left(\begin{array}{c}\tilde{G}_\varphi
\\ \tilde{G}_\chi \end{array}\right) , \\ \nonumber\zeta^{1/2}&=&
1-\frac{\tilde{G}_\varphi\tilde{G}_\chi}{8\pi^2\left(\tilde\Lambda^2-\tilde{m}^2\right)}
\log\frac{\kappa^2}{\tilde\Lambda^2}
\left(\begin{array}{cc}0 &
\tilde\Lambda^2-6\tilde{M}^2\\-\tilde{m}^2-6\tilde{M}^2&
0\end{array}\right).
\end{eqnarray}
The appropriate $\tilde\beta$ functions
\begin{equation}\label{app22}
\tilde\beta_\varphi=\frac{5\tilde{G}_\varphi^3}{16\pi^2},\quad
\tilde\beta_\chi=\frac{1}{16\pi^2}\left(2\tilde{G}_\chi^3+3\tilde{G}_\chi\tilde{G}_\varphi^2\right)
\end{equation}
contain no terms connected with mixing between light and heavy
scalars. So, the fourth term in Eq. (\ref{app20}) is removed, and
the RG relation for the $S$-matrix element becomes

\begin{equation}\label{app23}
{\cal D}\left(S_\mathrm{1PR}+S_\mathrm{box}\right)
=\frac{\partial\tilde{S}_\mathrm{1PR}^{(1)}}{\partial\log\kappa}
+{\tilde{\cal D}}^{(1)}\tilde{S}_\mathrm{1PR}^{(0)}=0,
\end{equation}
\begin{equation}\label{app24}
\tilde{S}_\mathrm{1PR}^{(0)}=\left(
\frac{\tilde{G}_\varphi^2}{s-\tilde{m}^2} -\tilde\alpha
+O\left(\frac{s^2}{\tilde\Lambda^4}\right)\right)
\bar{\tilde\psi}\tilde\psi\times\bar{\tilde\psi}\tilde\psi,
\end{equation}
\begin{eqnarray}\label{app25}
\frac{\partial {\tilde
S}_\mathrm{1PR}^{(1)}}{\partial\log\kappa}&=&
\frac{\bar{\tilde\psi}\tilde\psi\times\bar{\tilde\psi}\tilde\psi}{4\pi^2}
\left(-\frac{\tilde{G}_\varphi^4}{s-\tilde{m}^2}
\right.\nonumber\\&&\left.
+\frac{\tilde{G}_\varphi^2\left(-6\tilde\lambda\tilde{m}^2+\tilde{G}_\varphi^2\left(6\tilde{M}^2-s\right)\right)}
{\left(s-\tilde{m}^2\right)^2} \right.\nonumber\\&&\left.
+\tilde\alpha\tilde{G}_\varphi^2
+O\left(\frac{s^2}{\tilde\Lambda^4}\right)\right).
\end{eqnarray}

At $\tilde\alpha=0$ Eq. (\ref{app23}) is just the RG identity for
the scattering amplitude calculated in the absence of the heavy
particles. The terms of order $\tilde\alpha$ describe the RG
relation for the effective low-energy four-fermion interaction in
the decoupling region. The last one can be reduced in the RG
relation for the vertex describing the scattering of the light
particle (fermion) on the external field $\sqrt{\tilde\alpha}$
substituting the virtual heavy scalar:
\begin{equation}\label{app26}
{\cal D}\left(\sqrt{\tilde\alpha}\bar{\tilde\psi}\tilde\psi\right)
=\frac{\tilde{G}_\varphi^2}{8\pi^2}\sqrt{\tilde\alpha}\bar{\tilde\psi}\tilde\psi
+{\tilde{\cal D}}^{(1)}
\left(\sqrt{\tilde\alpha}\bar{\tilde\psi}\tilde\psi\right) =0,
\end{equation}
where
\begin{eqnarray}\label{app27}
&&{\tilde{\cal D}}^{(1)}=
\tilde\beta_\varphi\frac{\partial}{\partial\tilde{G}_\varphi}
-\tilde\gamma_\alpha\frac{\partial}{\partial\log\tilde\alpha}
-\tilde\gamma_m\frac{\partial}{\partial\log\tilde{m}^2}
-\tilde\gamma_\psi\frac{\partial}{\partial\log\tilde\psi},
\nonumber\\&& \tilde\gamma_\alpha=-{\cal
D}\tilde\alpha=-\frac{1}{8\pi^2}\left(3\tilde{G}_\varphi^2+O\left(\tilde\alpha\right)\right).
\end{eqnarray}

Eqs. (\ref{app23})-(\ref{app27}) is the main result of our
investigation. One can derive them with only the knowledge about
the low-energy couplings of heavy particle (\ref{app13}) and the
Lagrangian of the model with no heavy particles. One also has to
ignore all the heavy particle loop contributions to the RG
relation and the one-loop mixing between the heavy and the light
fields. Eqs. (\ref{app23})-(\ref{app27}) depend on the effective
low-energy parameters, only. But as the difference between the
original set of parameters and the low-energy one is of one-loop
order, one may freely substitute them in Eqs.
(\ref{app23})-(\ref{app26}). It is also possible to reduce the RG
relation for scattering amplitudes to the one for vertex
describing the scattering of light particles on the `external'
field induced by the heavy virtual particle. In fact, this result
is independent on the specific features of the considered model.

\end{document}